\documentclass[11pt]{article}
\usepackage{amsmath,amssymb,mathtools}
\usepackage{graphicx}
\usepackage{bm,color}
\usepackage[dvipsnames]{xcolor}
\usepackage[bookmarks=false,colorlinks=true,urlcolor=blue,citecolor=blue,linkcolor=blue]{hyperref}
\usepackage[letterpaper,
            twoside=true,
            bindingoffset=0in,
            left=1in,
            right=1in,
            bottom=1in,
            footskip=.25in]{geometry}
\usepackage{multicol}
\usepackage{setspace}
\usepackage{titling}
\usepackage{wrapfig}
\usepackage{soul}
\usepackage{floatrow}
\usepackage{caption}
\usepackage{titlesec}
\usepackage{fancyhdr}
\usepackage{mathptmx}  
\usepackage{bibentry}
\usepackage{enumitem}
\usepackage{authblk}
\usepackage[title]{appendix}

\title{Quantum Sensors for High Energy Physics}

\author[1]{Aaron Chou}
\affil[1]{Fermi National Accelerator Laboratory, Batavia, Illinois 60510, USA}

\author[2,3]{Kent Irwin}
\affil[2]{Department of Physics, Stanford University, 382 Via Pueblo Mall, Stanford, CA, 94305, USA}
\affil[3]{SLAC National Laboratory, 2575 Sand Hill Rd., Menlo Park, CA 94025, USA}

\author[4,5]{Reina H. Maruyama}
\affil[4]{Department of Physics, Yale University, New Haven, Connecticut 06520, USA}
\affil[5]{Wright Laboratory, Department of Physics, Yale University, New Haven, Connecticut 06520, USA}

\author[4]{Oliver K. Baker}
\author[3]{Chelsea Bartram}
\author[6]{Karl K. Berggren}
\affil[6]{Research Laboratory of Electronics, Massachusetts Institute of Technology, Cambridge, MA 02139, USA}
\author[1]{Gustavo Cancelo}
\author[7]{Daniel Carney}
\affil[7]{Lawrence Berkeley National Laboratory, 1 Cyclotron Road, Berkeley, CA 94720, USA}
\author[8,9,10]{Clarence L. Chang}
\affil[8]{Department of Astronomy and Astrophysics, University of Chicago, Chicago, IL 60637, USA}
\affil[9]{Kavli Institute for Cosmological Physics, University of Chicago, 5640 South Ellis Ave., Chicago, IL 60637, USA}
\affil[10]{Argonne National Laboratory, Lemont, IL 60439, USA}
\author[3]{Hsiao-Mei Cho}
\author[7]{Maurice Garcia-Sciveres}
\author[2]{Peter W. Graham}
\author[10]{Salman Habib}
\author[1]{Roni Harnik}
\author[4]{J. G. E. Harris}
\author[11]{Scott A. Hertel}
\affil[11]{University of Massachusetts, Amherst Center for Fundamental
Interactions and Department of Physics, Amherst, MA 01003-9337 USA}
\author[12]{David B. Hume}
\affil[12]{Time and Frequency Division, National Institute of Standards and Technology, Boulder, CO, USA}
\author[13,1]{Rakshya Khatiwada}
\affil[13]{Illinois Institute of Technology, Chicago, Illinois 60616, USA}
\author[14]{Timothy L. Kovachy}
\affil[14]{Department of Physics and Astronomy and Center for Fundamental Physics, Northwestern University, Evanston, IL 60208, USA}
\author[3]{Noah Kurinsky}
\author[4,5]{Steve K. Lamoreaux}
\author[15,16]{Konrad W. Lehnert}
\affil[15]{JILA, National Institute of Standards and Technology and the University of Colorado, Boulder, Colorado 80309, USA}
\affil[16]{Department of Physics, University of Colorado, Boulder, Colorado 80309, USA}
\author[17]{David R. Leibrandt}
\affil[17]{Department of Physics and Astronomy, University of California, Los Angeles, CA, USA}
\author[3]{Dale Li}
\author[18]{Ben Loer}
\affil[18]{Pacific Northwest National Laboratory, Richland, WA 99352, USA}
\author[19]{Juli\'{a}n Mart\'{i}nez-Rinc\'{o}n}
\affil[19]{Brookhaven National Laboratory, Upton NY 11973, USA}
\author[20]{Lee McCuller}
\affil[20]{Division of Physics, Mathematics and Astronomy, California Institute of Technology, Pasadena, CA 91125, USA}
\author[4,5]{David C. Moore}
\author[21,7]{Holger Mueller}
\affil[21]{Department of Physics, University of California, Berkeley, USA}
\author[1]{Cristian Pena}
\author[22]{Raphael C. Pooser}
\affil[22]{Quantum Information Science Section, Oak Ridge National Laboratory, Oak Ridge, TN 37831, USA}
\author[21]{Matt Pyle}
\author[23]{Surjeet Rajendran}
\affil[23]{Department of Physics and Astronomy, Johns Hopkins University, Baltimore, MD 21218, USA}
\author[24,25]{Marianna S. Safronova}
\affil[24]{Department of Physics and Astronomy, University of Delaware, DE, USA}
\affil[25]{Joint Quantum Institute, National Institute of Standards and Technology and the University of Maryland, College Park,
MD 20742, USA}
\author[2,3]{David I. Schuster}
\author[26]{Matthew D. Shaw}
\affil[26]{Jet Propulsion Laboratory, California Institute of Technology, Pasadena, CA 91109, USA}
\author[20]{Maria Spiropulu}
\affil{Division of Physics, Mathematics and Astronomy, California Institute of Technology, Pasadena, CA 91125, USA}
\author[19]{Paul Stankus}
\author[27]{Alexander O. Sushkov}
\affil[27]{Department of Physics, Boston University, Boston, MA 02215, USA}
\author[28]{Lindley Winslow}
\affil[28]{Massachusetts Institute of Technology, Cambridge, MA 02139, USA}
\author[1]{Si Xie}
\author[20]{Kathryn M. Zurek}

\date{\today}

\begin{document}

\maketitle
\tableofcontents

\section{Introduction}
A workshop was held at the Yale Quantum Institute in New Haven, CT on April 26-28, 2023 to identify enabling Quantum Information Science technologies that could be utilized in future particle physics experiments targeting High Energy Physics (HEP) science goals.   Following the charge copied in Appendix~\ref{s:charge}, this workshop aimed to build upon
previous workshops and planning exercises, including the Quantum Sensing for HEP
workshop at Argonne National Laboratory in December 2017~\cite{ahmed2018quantum}, the HEP Instrumentation Basic
Research Needs workshop in December 2019~\cite{DetectorBRN2020}, and the HEP Quantum Information Science Enabled Discovery (QuantISED) sensors workshop in
May 2020~\cite{barry2021opportunities} in order to 
``identify particularly promising approaches in the domain of quantum
sensing that can be utilized by future HEP applications to further the scientific goals of the
community as outlined by the (2014) P5 Report~\cite{P5report2014} and in the recent Snowmass 2021 Report~\cite{osti_1922503}."
The organizing committee included the two Snowmass 2021 Instrumentation Frontier conveners for quantum sensors.  Participants were invited based on their ongoing activities in the area of quantum sensors as applied to fundamental physics topics.  In consultation with DOE national lab points of contact for quantum sensing, this cohort was identified through recent awards funded by the DOE QuantISED program described in appendix~\ref{a:quantised}, quantum sensing R\&D roles in the National Quantum Initiative Science Research Centers (NQISRC), participation in recent quantum sensors workshops including activities in Snowmass 2021, and publication of recent papers relevant to this topic.  In particular, representatives from each of the ongoing QuantISED efforts were invited to participate in this workshop
to discuss the status and future directions for the quantum
sensing technologies that they are currently exploring, targeting new capabilities for HEP science. In addition, while not the main focus of the National
Quantum Initiative, several of the NQISRCs contain quantum sensing programs which develop sensor technologies
which are relevant to HEP applications. Representatives from these NQISRCs were also invited to join the workshop to better engage with these ongoing research programs.
In order to obtain a broad cross-section of U.S. quantum sensing efforts, the workshop participants included a mix of DOE national laboratory staff and university faculty, as well as researchers from NIST and NASA labs.

Following the charge, the focus of the workshop was on opportunities in the development and deployment of quantum sensor hardware, with the understanding that the significant 
other components of ongoing quantum information science (QIS) research
would be covered in separate studies.  To keep the small workshop manageable, representatives from the various experimental quantum sensor R\&D efforts were invited as well as a group of theorists who have been identifying HEP use cases for this hardware.  Junior faculty-level early career scientists working in areas utilizing quantum sensors were encouraged to participate. 
New experimental concepts utilizing QIS techniques were also solicited on the Snowmass mailing lists and Slack channels where members of the community were invited to give short hybrid talks in the workshop town hall.  Short summaries of the talks are documented in appendix~\ref{a:townhall} and the presentation slides are available on the workshop indico site \href{https://indico.fnal.gov/event/59102/}{https://indico.fnal.gov/event/59102/}.

To gather information on progress and current ideas on quantum sensing for HEP science, workshop participants were asked to form small groups with common science and technology interests and discuss responses to a series of salient questions listed in Appendix~\ref{s:questions} concerning each identified intersection of HEP science with quantum sensing technology shown in figure~\ref{fig:table_science_tech}.  The findings from each working group were then presented in a plenary session including all participants, and are summarized in this report.  

\section{Executive summary}
Strong motivation for investing in quantum sensing arises from the need to investigate phenomena that are very weakly coupled to the matter and fields well described by the Standard Model. These can be related to the problems of dark matter, dark sectors not necessarily related to dark matter (for example sterile neutrinos), dark energy and gravity, fundamental constants, and problems with the Standard Model itself including the Strong CP problem in QCD. Resulting experimental needs typically involve the measurement of very low energy impulses or low power periodic signals that are normally buried under large backgrounds. 

With a focus on HEP applications, we propose to adopt a broad definition of the term quantum sensor as any new quantum device or technique that has the potential to achieve greater reach towards beyond-the-standard-model physics than that achievable through conventional techniques traditionally used in HEP.  Some definitions used elsewhere such as ``Quantum 2.0'' as ``a class of devices that actively create, manipulate and read out quantum states of matter, often using the quantum effects of superposition and entanglement'' highlight an important class of these techniques. These include sensors that evade ``Standard Quantum Limits'' (SQLs) that are imposed by classical sensors. However, while Quantum 2.0 devices have an important role in HEP applications in the future, 
focusing only on this category is viewed as too constraining of the future HEP quantum program, which should more broadly explore new experimental quantum techniques and sensing strategies, in particular newly maturing technologies which have leveraged broader investment in R\&D from the new QIS programs and from other non-HEP sources.

Quantum sensors includes qubits and continuous-variables quantum devices of various types as developed by the quantum computing community, quantum materials and spin ensembles developed by the condensed matter and materials science community, and atomic and optomechanical sensors, clocks, and interferometers developed by the AMO and gravitational wave communities, in each case repurposed and re-optimized to target HEP science.  However, new quantum sensing techniques could also be enabled by HEP technologies or other ultrasensitive technologies not commonly viewed as Quantum 2.0.  For example, extremely high quantum efficiencies of transition-edge sensors can be used for the measurement of entanglement generation while the low dark rates of SNSPDs and the ultra-low noise achieved by repeated CCD readout provide previously unimaginable new capabilities in high fidelity single quantum detection.  Such technologies are already being deployed in groundbreaking non-HEP applications such as quantum imaging and microscopy, optical quantum computing, and quantum networks which may themselves be used as resources for future HEP experiments.  SRF cavity technology and superconducting expertise that was developed over decades for accelerators is now being brought into the quantum regime by the Superconducting Quantum Materials and Systems (SQMS) center, one of the new DOE quantum centers, in order to develop quantum processors and sensors for HEP.  HEP instrumentation expertise may also be used to economically field large arrays of quantum sensors as may be necessary to achieve large active area coverage or to search for extremely long wavelength phenomena.  The high speed and precision timing of microscopic or mesoscopic quantum devices may also be essential to future accelerator-based or collider experiments with high event rates in order to image the overlapping scattering events which would not be individually resolvable with conventional sensors.

For this report, we focus on new and emerging quantum sensing technologies and methodologies which have not been traditionally used in HEP experiments and whose development could efficiently leverage the large investments being made both in the U.S. and world-wide in quantum science.  Through this strategy of cross-disciplinary engagement, we hope to bring new capabilities to the field of experimental HEP and usher in the next era of scientific discovery.

The need for similar quantum sensing technology is shared by many disparate parts of the scientific community. The current approach of ``stove-piped" funding of the same technology across different disciplines (HEP funded by DOE and NSF, precision measurements funded by NIST, astronomical sensors funded by NASA, quantum computing funded by DOD) presents a barrier to rapidly make the necessary progress.  To be sure, many of the quantum sensing technologies will continue to improve in performance due to ongoing U.S. investment in QIS and other fields, particularly in the non-HEP science contexts in which the technologies had originally been developed.  However, cross-disciplinary engagement will be necessary to adapt these technologies to achieve performance meeting HEP experimental requirements.  These requirements are often more stringent than those of the original application and the operating parameters can also be completely different from what was originally envisioned.  For example, quantum computing researchers do not typically design their qubits to work within the high magnetic fields which are needed for some HEP detection applications.  

Furthermore, much work will be necessary to deploy and commission quantum sensing devices in real-world HEP experiment conditions which are often not the pristine, clean, field-free environments enjoyed by quantum computing test stands.  Early strategies to simply purchase the quantum devices from research groups funded outside of HEP were not successful as these quantum detectors are not turn-key devices operable by non-experts.  Rather, just like HEP detectors, they comprise complex systems in which to preserve and control signal information stored as single quanta or in similarly tiny amounts of energy -- every aspect of the environmental operating conditions and the low-noise control and readout electronics must be carefully designed and evaluated in extended commissioning and operations phases.  This complexity thus requires full scientific engagement of QIS scientists and engineers in neighboring fields to participate in a cross-disciplinary co-design process in which actual performance data from prototype devices is fed back into the next iteration of device design and fabrication until the specifications for the HEP experiment can be met.

\subsection{Key Findings}
\begin{itemize}
\item Large areas of well-motivated HEP model parameter space remain unexplored due to limitations in detector technologies traditionally employed in HEP.
\item Many mature quantum sensing technologies have been developed in neighboring fields and can be adapted to larger-scale HEP applications.
\item Other quantum sensing techniques hold great promise but still need further development to achieve the technical requirements for HEP science targets.
\item In order to harness the full potential of quantum sensing, a robust HEP quantum program should therefore include a diverse portfolio of detector R\&D, pathfinder experiments, and full-scale experiments. The portfolio should also include support for theory which has played a crucial role in recognizing the opportunities provided by quantum sensing for HEP applications. 
\item Scientific and technical collaboration with researchers in neighboring disciplines, including those funded outside of HEP, provide crucial access to previous experience and ongoing innovation in broader QIS\&T research 
\item Strategic utilization of new or existing foundry facilities to fabricate quantum sensors for HEP science projects will be a necessary element of the future HEP quantum sensors program.  
\end{itemize}

\subsection{Strategy to optimize HEP program utilization of QIS technology for sensing} \label{Sec:Strategy}

\begin{enumerate}
\item	Ensure stable funding as a crucial element of workforce development. Enable groups to retain the personnel expertise they have managed to build up and to hire where required. Especially in the QIS field, which has extremely strong demand for expertise outside HEP, it is very difficult to retain full time staff or hire competitively without the stability provided by base funding.  
\item	Explore a strategy for supporting and operating quantum device fabrication facilities to provide critical capabilities for the HEP program. Consider a quantum device fabrication exchange to connect the foundry and assembly capabilities among the different efforts in the HEP QIS community and avoid gaps limiting the implementation of new quantum sensor technologies for the program. 
\item	Provide funding for theory support for quantum sensing/metrology.  Most QIS-for-HEP experimental techniques were originally proposed by HEP theorists.
\item	Advance QIS detectors from proof-of-concept to science.
\begin{enumerate}
\item	HEP’s core science questions have a demonstrated need for new capabilities that new quantum techniques are poised to provide. There is potential for quantum technologies to dramatically impact HEP science well before the goals of quantum computing are realized. The dream is that quantum sensors can now enable breakthrough science, if adapted and scaled up as needed to address HEP science targets.  A exposition of ongoing technology challenges, needs, and gaps is provided in section~\ref{s:challenges}.
\item	A better “impedance match” is needed between the impressive capabilities being developed in the QIS program to the critical HEP science targets.  Important elements to consider for the HEP experimental program include:
\begin{itemize}	
\item	Leveraging expertise of HEP scientists and engineers who are experienced with “big science”
\item	Providing support on timescales that allow multiple iterations of data-taking and improvement
\item	Establishing an infrastructure of collaboration and utilizing high-performance computing infrastructure that exists within the HEP community
\item	Funding university scientists and engineers working at the intersection of QIS with HEP, in addition to national lab personnel, following the model of other DOE programs.  This would help to develop the collaborations needed for larger-scale projects. 
\item	Establishing experimental program support at national labs.  Incorporating QIS ``quantum enhancement" into extended HEP search campaigns poses integration and operations challenges that will need continuity and expertise to be maintained via professional staffing.
\end{itemize}
\end{enumerate}
\end{enumerate}

\section{Science drivers for quantum sensors}
\label{Sec:ScienceDrivers}
The workshop was organized following the blueprint of a Basic Research Needs study in which the relevant HEP science drivers were first identified along with technology opportunities and gaps that prevent further progress in each research direction.  This section will summarize the collection of science topics which would benefit from the new sensing capabilities afforded by QIS technologies.  The next section will go into more detail on the specific technologies currently being considered by the HEP and cross-disciplinary quantum research community.

\subsection{Dark waves}
If the mass of dark matter is much less than $\sim10$\,eV, then the known density of dark matter in the galaxy requires that there are many particles per quantum state. Thus, low-mass dark matter behaves more like a wave than a particle. These candidates behave similarly to radio waves, which have many photons per quantum state. The most well-known and the best-motivated wave dark matter candidate is the QCD axion, which solves the strong CP problem via its electric dipole moment (EDM) interaction. There are other wave dark matter candidates including hidden photons and scalar particles. Unlike searches for heavier dark matter, in which individual dark matter particle scattering events are detected, searches for these dark matter waves require quantum radio techniques. Dark waves interact with the standard model in multiple ways, coupling to photons, or to nuclear spins, or by causing apparent fluctuations when measuring physical constants. While classical radio technology is quite mature, the power from dark waves is very low, and there is a tremendous new opportunity in deploying new quantum technologies for ultra-sensitive measurement of dark waves~\cite{Sushkov2023c}. Coverage of most of the QCD axion parameter space is within reach, using the experimental technologies described below. Achieving this goal will require solving a number of challenges, such as adapting quantum sensors to operate in or near the high magnetic field environments needed for some of these experiments.

While dark waves interact with the standard model through photons, nuclear spins, or variations in physical constants, in all cases these interactions have very low signal power, requiring the implementation of new quantum sensing techniques. At higher frequencies (above $f\sim 1$\,GHz), interaction with electromagnetism can be sensed with single-photon sensors (including qubits, Rydberg atoms, or pair-breaking sensors). At lower frequencies,  thermal noise is significant, corresponding to  $k_B T > h f$, which occurs below $f \sim 300$\,MHz in a dilution refrigerator. The implementation of single-photon sensors and the creation and preservation of photon-number states (Fock states) can be prohibitively challenging at these frequencies, and coherent continuous-variable quantum sensors, including radio-frequency quantum upconverters (RQUs) and Josephson parametric amplifier (JPA) squeezers, can be used. Dark-wave interaction with nuclear spins can be sensed by using the nuclear spins themselves as quantum sensors. Atomic clocks and neutral-atom interferometers are used to measure scalar interactions in variations in physical constants. The QuantISED program includes active R\&D on quantum electromagnetic sensors and nuclear spin sensors for dark waves.

\subsection{Dark particles}
Traditional particle dark matter searches operate in the billiard-ball regime where a dark matter particle scatters off an individual nucleus or electron, transferring enough momentum that the atomic binding of that nucleus or electron is negligible. This results in an ionization signal, measured with traditional methods. Lower mass dark matter particle interactions, below ionization threshold and even below semiconductor bandgaps, cannot couple to single atoms but instead to coherent modes in crystals~\cite{Knapen:2017ekk}, superconductors~\cite{Hochberg:2015pha}, or superfluid He~\cite{Schutz:2016tid}, which have much lower energy excitations. These excitations can be phonons, rotons, magnons~\cite{Trickle:2019ovy}, and quasiparticles in superconductors. Typically, phonons, rotons, magnons are collected on superconductors and turned into quasiparticles. Therefore, much of the quantum sensor program for low mass particle dark matter focuses on the measurement of quasiparticles (pairbreaking sensors - aka broken Copper pairs) in superconductors, typically Al as the superconductors of choice, with a Cooper pair breaking energy of $\mathcal{O}$(1~meV).

The electrical properties of a superconductor depend on the density of Cooper pairs (superconducting electrons) and quasiparticles (normal electrons) in the superconductor. The presence of Cooper pairs is measured through the electrical resistance of a superconductor in its transition by a Transition Edge Sensor (TES). The quasiparticle density is measured by the variation in inductance of the superconductor by a Kinetic Inductance Device (KID). The quasiparticle density can also be measured by the change in the capacitance of a Cooper pair box coupled to the superconductor in a Quantum Capacitance Detector (QCD), using quasiparticle traps as in the Superconducting Quasiparticle-Amplifying Transmon (SQUAT)~\cite{fink2023superconducting}.
 These devices can all be operated in a linear regime: the electrical response is proportional to the number of quasiparticles and therefore to the deposited energy.  There are also fundamentally non-linear devices -- notably SNSPDs. The best energy resolution is currently achieved by TESs, MKIDs have the ability to multiplex a large number of channels, QCDs have demonstrated single THz photon counting, and SNSPDs operate with nearly zero dark counts at $>$1 K temperatures. None of these devices are near fundamental limits of sensitivity. Ongoing development is mainly concerned with understanding and improving the practical  issues that currently limit device sensitivity, with scaling to systems suitable for dark matter searches. The practical issues include Materials Science problems, electrical system aspects such as control and readout, and manufacturing issues such as thin film uniformity.

\subsection{Cosmology, dark energy, phase transitions}
The top cosmological science priorities for HEP are to understand the physics of inflation, dark energy, and dark matter, with cosmological surveys, such as CMB-S4, DESI, and Rubin Observatory's LSST, providing the needed observational facilities. Future facilities including near-IR spectroscopic surveys, Line Intensity Mapping surveys (in both radio and mm-wave), and sub-mm observatories will drive this science forward through measurements of new spatial and temporal scales. Notably, the technology development required for all of these future surveys involves scalable high-efficiency sensing of photons with longer than optical wavelengths, an area that shares broad overlap with QIS. Interestingly, ambitious applications of QIS may enable new approaches to cosmological studies that complement these major surveys such as enabling new astrophysical observables and direct terrestrial measurements.

\subsubsection{Improved sky observations for HEP science}
\label{subsubsec:improved_sky}

Quantum-engineered sensors and sensor arrays can greatly improve observations of astronomical objects; in particular we highlight quantum improved interferometry (see Figure~\ref{fig:Quantum_Netowrk}), both single-photon \cite{Gottesman_2012,brown2023interferometric} and two-photon \cite{Stankus_2022, crawford2023quantum}, for precision astrometry in the post-GAIA era.  Applications with the potential to directly HEP science include \cite{Stankus_2022, derevianko2022quantum, Fedderke_2022gw}:

\begin{itemize}
    \item Precision measurement of the orbits of close, spectroscopic binaries provides an absolute distance measure which is completely independent of parallax.  This could provide a great systematic improvement in the first rung of the distance ladder and directly address the Hubble tension

    \item Dynamic astrometry of wide binaries and stellar clusters can constrain the properties of dark matter in the Galaxy, and also provide tests of alternate theories of gravity.

    \item Precision astrometry, and interferometric imaging, of a gravitational microlensing event in progress can provide detailed tests of GR, as well as information on dark matter candidates.

    \item Precision astrometry of curated bright objects can provide a way to detect the presence of gravitational waves at low frequencies, in the micro-Hertz regime.  These would provide cosmologically relevant information on structure formation and galaxy evolution through constraints on supermassive black hole mergers, as well as being sensitive to possible phase transitions in the early thermal Universe \cite{Fedderke_2022gw}.
\end{itemize}

These kinds of observational programs would be an opportunity for Cosmic Frontier investigation to expand beyond sky survey experiments and focus on specific observables tailored to HEP science targets.

\subsubsection{Direct Detection of Dark Energy}

Traditional probes of the equation of state of dark energy (the determination of $w$ and $w_a$) are performed using cosmic surveys. These surveys are gravitational probes of the equation of state of dark energy and while important, they have a limited ability to probe this equation of state when the equation of state gets close to (but different from) -1. There is however the exciting possibility that the dynamical component of dark energy (the component that is different from a cosmological constant) could be explored in the laboratory, leading to significantly more potent probes of the equation of state. This possibility mirrors the situation in dark matter -- gravitational probes of the equation of state of dark matter are able to place rather weak limits on the scattering of dark matter with baryons. However, laboratory probes of this scattering are orders of magnitude better than the gravitational constraints.

There are two distinct classes of laboratory signatures of dark energy. Interestingly, both of these classes of signatures can be optimally probed using a variety of quantum sensors.  First, the coherent slow roll of the dark energy would appear effectively as a Lorentz violating background in the laboratory \cite{Pospelov:2004fj, Graham:2020kai}. This Lorentz violating background can induce precession of nucleon/electron spins. It can also cause rotations in the polarization of light. The former class of signatures can be looked for using a variety of spin precession experiments that are optimized to look for the low frequency dc precession caused by the dark energy. The latter class of signatures can be probed using measurements of the polarization rotation of the cosmic microwave background \cite{Bean:2005ru,Fedderke:2019ajk}. While these two signatures are typical of axion-like models of dark energy where a shift symmetry protects the mass of the dark energy field (and thus keeps its low mass technically natural), it is possible that the dark energy field may manifest naturalness through more complex manifestations of symmetry \cite{Hook:2019mrd}. If so, the slow roll of the dark energy field would also induce shifts in atomic or nuclear clock frequencies and these may be detectable in AMO experiments \cite{Brzeminski:2022sde, Brzeminski:2020uhm}. 

The search for this class of dark energy signals leverages QIS HEP investments in the search for ultralight dark matter. In comparison to dark matter, the energy density in the dark energy in the galaxy is about five orders of magnitude less than the dark matter energy density since the dark energy, unlike the dark matter, does not cluster into galaxies. However, since the above class of experiments measures the amplitude of the dark energy field rather than the power in it, the dark energy signal is only about 2 - 3 orders of magnitude weaker than a comparable dark matter signal. But, the dark energy signal, while weaker, also has some advantages. It is essentially coherent over the age of the universe unlike the dark matter signal which is coherent only for a million periods of the dark matter mass. Thus, the dark energy signal can be looked for by integrating for long periods of time, if technological levers that can operate at low (or dc) frequencies can be suitably developed. 

Second, the coherent slow roll of the dark energy field can be converted into various forms of dark radiation \cite{Berghaus:2020ekh}. Since this dark radiation is produced due to the decay of the dark energy in the recent universe, it is unconstrained by CMB probes of light degrees of freedom. The energy density in this dark radiation can be as large as (meV)$^4$, about 4 orders of magnitude larger than the energy density in the CMB. Experiments to directly dark energy radiation would resemble cosmic microwave background experiments, using large arrays of single mode quantum sensors to gather statistics of the dark blackbody. The detection of such dark radiation, with an energy density substantially bigger than that of the CMB, would uniquely point to dark energy being the source of such radiation. 

\subsubsection{Inflation and gravitational waves}

Phase transitions occurring in the early universe can be probed via the characteristic spectrum of stochastic gravitational waves that they produce.  The usual example is the flat gravitational wave background sourced by horizon-size quantum fluctuations during the epoch of cosmic inflation, which are imprinted on the cosmic microwave background as B-mode polarization patterns.  While direct detection of these inflationary gravitational waves remains a long-term challenge, other possible post-inflationary cosmological phase transitions may produce signals more amenable to discovery via the guaranteed universal coupling to gravity of any new dark sector. Spectral features of gravitational waves from such phase transitions can be used to directly probe the number of degrees of freedom and the beta function of physics coupled to the standard model all the way to the high energies of the phase transition \cite{Brzeminski:2022haa}. Moreover, recent results from a wide variety of pulsar timing arrays \cite{Agazie:2023eig, NANOGrav:2023wsz} show that the universe is awash with stochastic gravitational waves at nano hertz frequencies. While the origin of these gravitational waves is unknown, the blue tilt in the spectrum hints at the possibility that these gravitational waves may arise from physics beyond the standard model. The discoveries at Nanograv and LIGO suggest that the gravitational wave sky is rich with a number of opportunities for discovery. As the potential for discoveries in this arena are better digested by the community, it is likely that there would be increased excitement for broad probes of the gravitational wave spectrum. Investments made in QIS HEP can potentially be used to provide new sensitivity to gravitational waves over a wide range of frequencies. For example, quantum sensing technologies built for a wide variety of dark matter experiments such as atom and optical interferometers as well as electromagnetic resonator experiments can be deployed for gravitational wave detection in a number of frequency bands.

\subsection{Quantum gravity}
One of the greatest outstanding issues in physics is a quantum description of gravity. General relativity, as a purely classical theory, can be considered as part of the standard model. Its quantum formulation is then BSM physics that we must strive to theoretically understand and experimentally test.

Theoretically, quantum gravity is a major challenge due to formal incompatibilities between general relativity and quantum field theory. Experimentally, it is extremely challenging due to the weak interaction strength of gravitation and the fragility of macroscopic quantum states. There are two major avenues to observe quantum gravity effects. The first is in detecting gravitational interactions of quantum matter. This is deeply related to macroscopic tests of quantum mechanics, as weak gravitational interactions are only likely to be observed in large systems. 

The second is determining observational consequences of a quantum micro-physical description of gravity that manifest at detectable length and energy scales. One such emerging theory predicts metric fluctuations arising from an underlying entanglement-entropy description of gravitation~\cite{liPRD23InterferometerResponse,verlindePLB21ObservationalSignatures,verlindePRD22ModularFluctuations,zurekPLB22VacuumFluctuations}, which implies observable gravitational noise from entropic fluctuations. Single-photon detectors from emerging QIS HEP efforts can help in converting traditional optical inteferometry into HEP-style particle counting experiments, to detect the signal energy of quantum gravity metric fluctuations deeply below the quantum imprecision limit imposed by photon shot noise~\cite{mcculler2022singlephoton, ngPRA16SpectrumAnalysis}. The conversion to this alternate photon-counting readout can immediately employ emerging low-background quantum detection technologies such as SNSPDs. It additionally motivates more sophisticated methods to reduce backgrounds, which can employ quantum memory technologies~\cite{gorshkovPRA07PhotonStorage} to implement low spectral leakage photon counting using customizable temporal apodization of optical fields. Furthermore, non-Gaussian readout of photon counting motivates non-Gaussian state preparation as a measure to further gain quantum enhancement beyond using (Guassian) squeezed states.  A notable advantage of optical interferometry as a quantum technology platform is that, in many circumstances, it does not require experiments to live within cryostats. Thus, it provides opportunities to develop and utilize quantum technologies translatable to general-use applications. Altogether, interferometric searches for quantum gravity motivate development in a host of emerging quantum technologies.

These probes of quantum gravity, which may potentially reveal the UV features of gravity, are distinct from attempts to simulate simple quantum systems and reinterpret them in the language of gravity. Such simulations would more appropriately belong in the portfolio of quantum simulation rather than quantum sensing. 

\subsection{Testing Quantum Mechanics}

Quantum mechanics is the bedrock of physics. But, its axioms were derived phenomenologically, leading to the enticing possibility that these axioms are an approximation to a more complete theory. There is a strong need to develop consistent deviations from these axioms and experimentally probe them. Linearity of time evolution is a central axiom of quantum mechanics. But, in every other known physical system, linearity is an approximation to a more complex system. Recent theory work \cite{Kaplan:2021qpv, Polchinski:1990py, Kibble:1978vm, Kibble:1979jn, Stamp:2015vxa} has shown that there are in fact a broad class of consistent (causal and gauge invariant) extensions of quantum mechanics where the time evolution of the quantum system is nonlinear. In this class of theories, one can view the time evolution equation of a quantum field theory as a power series expansion in the quantum states and the term that is linear in the quantum state is simply the first term in this expansion. Thus, linear quantum mechanics emerges as the leading order approximation of a more complex time evolution.

This broad class of theories is poorly constrained~\cite{Kaplan:2021qpv, Polchinski:1990py} by the large number of quantum mechanical tests that have been performed. However, they can be readily tested by a number of suitably optimized setups using experimental technologies that currently exist \cite{Polkovnikov:2022hkg, Broz:2022aea, Page:1981aj}. The combination of weak limits from existing experiments and the ability to devise potent probes of nonlinear quantum mechanics using presently available experimental technologies presents an interesting scientific opportunity. 

A broad suite of QIS HEP tools can be used to perform these tests. These tests feature a combination of precision sensing instruments such as voltmeters, magnetometers and accelerometers, deployed in concert with macroscopic classical sources of electromagnetism and gravitation such as high electric and magnetic fields as well as large test bodies. The technologies necessary to search for nonlinear quantum mechanics are thus similar to the kinds of technologies necessary to search for new fundamental interactions in the laboratory where such interactions are sourced by laboratory based objects. Investments in these areas can thus be leveraged to probe the fundamental nature of quantum mechanics. It is important to highlight a major technological opportunity that exists with such nonlinear modifications of quantum mechanics. The linearity of quantum mechanics constrains the possible applications and exploitability of quantum systems.

In addition to nonlinear modifications of quantum mechanics, it may also be that quantum mechanics is fundamentally not unitary. Indeed, we recognize that interactions of quantum systems with the environment leads to decoherence, in effect resulting in open quantum systems evolving in a non-unitary way. There is nothing logically inconsistent with non unitary time evolution. Why then should quantum evolution be fundamentally unitary when we see plenty of examples of systems where the time evolution is not unitary? 

Using the mathematical framework developed to describe decoherence in quantum mechanics, it is possible to write down theoretical extensions of quantum mechanical evolution where  decoherence is fundamental \cite{Weinberg:2016uml}. The existence of theoretical puzzles such as the black hole information problem which places the unitary nature of quantum mechanics in conflict with the causal structure imposed by General Relativity motivate the need to test the unitary nature of quantum time evolution. Tests of this fundamental property of quantum mechanics can also be performed using QIS tools such as atomic and ion clocks.

\subsection{BSM physics with accelerators and colliders}
Future proposed ultra high intensity accelerator and collider experiments such as the Future Circular Collider (FCC) or the Muon Collider will operate at intensities that are many orders of magnitude higher than the current state-of-the-art in order to achieve their physics and science goals.
These extreme conditions impose much more stringent requirements on the performance of particle detectors, particularly on position and time resolution.
For example, the FCC-hh is estimated to have a factor of 20 larger amount of pileup -- the number of simultaneously occurring collisions at the interaction point -- than the currently operating LHC. 
These ultra intense collider experiments will require detectors with a timing precision of 1 ps to assign particles to the correct interaction vertex, reduce beam induced backgrounds, and reconstruct the physics objects on an event by event basis, all necessary to achieve their ultimate science reach. 
Moreover, certain classes of proposed exotic particles including millicharged particles and low mass weakly interacting particles produce sub-eV energy depositions, presenting tremendous challenges for their efficient detection. 

Quantum sensors have the potential to contribute significantly to solving these demanding challenges. 
Their uniquely low energy detection thresholds have the ability to detect the aforementioned classes of BSM particles with sub-eV energy deposition, for which no other particle detection technology currently exists.
Exotic particles such as millicharged particles, whose ionization energy loss is highly suppressed by its tiny electric charge, or low mass weakly interacting particles, whose interaction with material results in sub-eV recoil energies, can be detected by quantum sensors.
Furthermore, the nano-scale structures characteristic of quantum sensors lends naturally to detectors with sub-micron position resolution capabilities.
Finally, the signal generation process in superconducting sensors such as SNSPDs operates at sub-ps scales, resulting in detectors with natural capability for picosecond time resolution.

Many of these capabilities have individually already been demonstrated.
For example, SNSPDs have demonstrated 2.7~ps time jitter for detection of a single photon and promise to provide next-generation collider experiments with the precision timing requirements needed as a single sensitive element.
A strong R\&D program to experimentally demonstrate and optimize the ultimate time resolution performance of SNSPDs to charged particles, expected to deposit larger energies compared to single photons, will pave the road towards understanding the full extent of the physics case for such quantum sensors. Additionally, SNSPDs are expected to be radiation hard up to the fluences expected for future collider experiments. However, experimental verification is necessary to prove that the sensor’s performance metrics are maintained after radiation exposure and should be an active area of research.

The combination of sub-eV energy detection threshold, sub-micron position resolution, and picosecond time resolution all realized in the same quantum sensor has the potential to revolutionize particle detection technology at future accelerator and collider experiments. 

\subsection{Fundamental Symmetries and Interactions}
Quantum sensors will also advance probes of violations of fundamental symmetries (such as C, P, T symmetries) and new interactions (such as forces and torques exerted on standard model sensors by laboratory test bodies). This includes the potential application of ultracold molecules to look for the electric dipole moment of nucleons and electrons, molecular spectroscopy to look for new short range interactions, atom interferometry to look for new forces and a variety of spin sensors to search for new torques. A number of these technologies can also be used to look for time varying signals caused by dark matter waves - thus, QIS technologies developed for dark matter can be leveraged to look for this class of signals as well. 

Observation of an electric dipole moment (EDM) in any experimental system (electron, neutron, proton, atom, molecule) at current or near-future sensitivity would yield exciting new physics. Assuming maximal breaking of CP symmetry, EDMs probe beyond the Standard Model (BSM) mass scales well beyond those directly probed at high energy colliders \cite{EDM}. 
A well coordinated program of complementary EDM searches in atomic/molecular, nuclear, and particle physics experiments is needed to discriminate among the many viable BSM theories, test baryogenesis scenarios for the observed matter-antimatter asymmetry of the Universe, or tell whether CP symmetry is spontaneously or explicitly broken in Nature.

Atoms and molecules are exceptionally  sensitive platforms for precision measurements of symmetry violations, including CP-violation (CPV) through EDMs~\cite{Safronova2018,Chupp:2017rkp}.  These experiments currently set the best limits on the electron EDM, semileptonic CPV interactions, and quark chromo-EDMs; they also are competitive with nEDM for sensitivity to quark EDMs and $\theta_{QCD}$~\cite{ACME2018,Graner2016}.  These searches  use atomic, molecular, and optical (AMO) techniques to achieve their goals, and 
leverage the very advanced quantum control  techniques \cite{EDM}.
The basic idea of such EDM searches is that CP-violating electromagnetic moments of atomic and molecular constituents, in particular electrons and nuclei, are amplified by the extreme internal electromagnetic environment. 

Atomic and molecular  CP-violation searches are already probing energy scales within the predictions of BSM theories.  Even more excitingly, AMO searches for fundamental symmetry violations offer realistic pathways to orders of magnitude of improvement in the not-too-distant future.  First, modern molecular experiments are still relatively new, and recent advances such as laser cooling and trapping at ultracold temperatures~\cite{Hutzler2020Review,Fitch2021Review}, matrix isolation, and advanced control in molecular ion traps offer orders of magnitude increases in sensitivity through improvements in coherence time and count rates of existing approaches \cite{EDM}.  Such experiments, directly exploring HEP-relevant science, are collaborative, larger scale, and beyond the usual 3-year timescale that can be readily supported by existing AMO programs. 

Second, heavy nuclei with octupole deformations can lead to enhanced sensitivity to hadronic CP-violation, up to a thousandfold larger than in spherical nuclei. Combined with enhancement from large molecular fields, molecular species with deformed nuclei can be up to $\sim\!10^7$ times more intrinsically sensitive~\cite{Flambaum2019Schiff} than the current state of the art~\cite{Graner2016}. 
As a result, radioactive atoms and molecules offer extreme nuclear charge, mass, and deformations, and may be worked with efficiently with the advanced quantum control toolset of AMO. These rare systems offer an unprecedented amplification of both parity- and time-reversal violating properties \cite{RM}.

Use of radioactive atoms and molecules requires extensive molecular spectroscopy, and in particular, dedicated facilities which can perform this spectroscopy with very short-lived species \cite{RM}. These developments are well aligned with the progress of radioactive beam facilities around the world, such as the DOE facility for rare isotope beams (FRIB), where unique access to large amounts of actinide nuclei will be provided. Thus it is critical to support the infrastructure needed to integrate table top experiments with radioactive beam facilities.

Atom and molecule-based searches for fundamental symmetry violations have advanced rapidly in recent
years, and have excellent prospects for further advances.
Improvements in sensitivity of one, two-three, and four-six orders of magnitude appear to be realistic on the few, 5--10, and 15--20 year time scales, respectively, by leveraging major advancements made using quantum science techniques and the increasing availability of exotic species with extreme sensitivity. These gains open an exciting pathway to probe PeV-scale physics using “tabletop” scale experiments \cite{EDM}.

\begin{figure}[ht!]
    \centerline{
    \includegraphics[width=0.95\columnwidth]{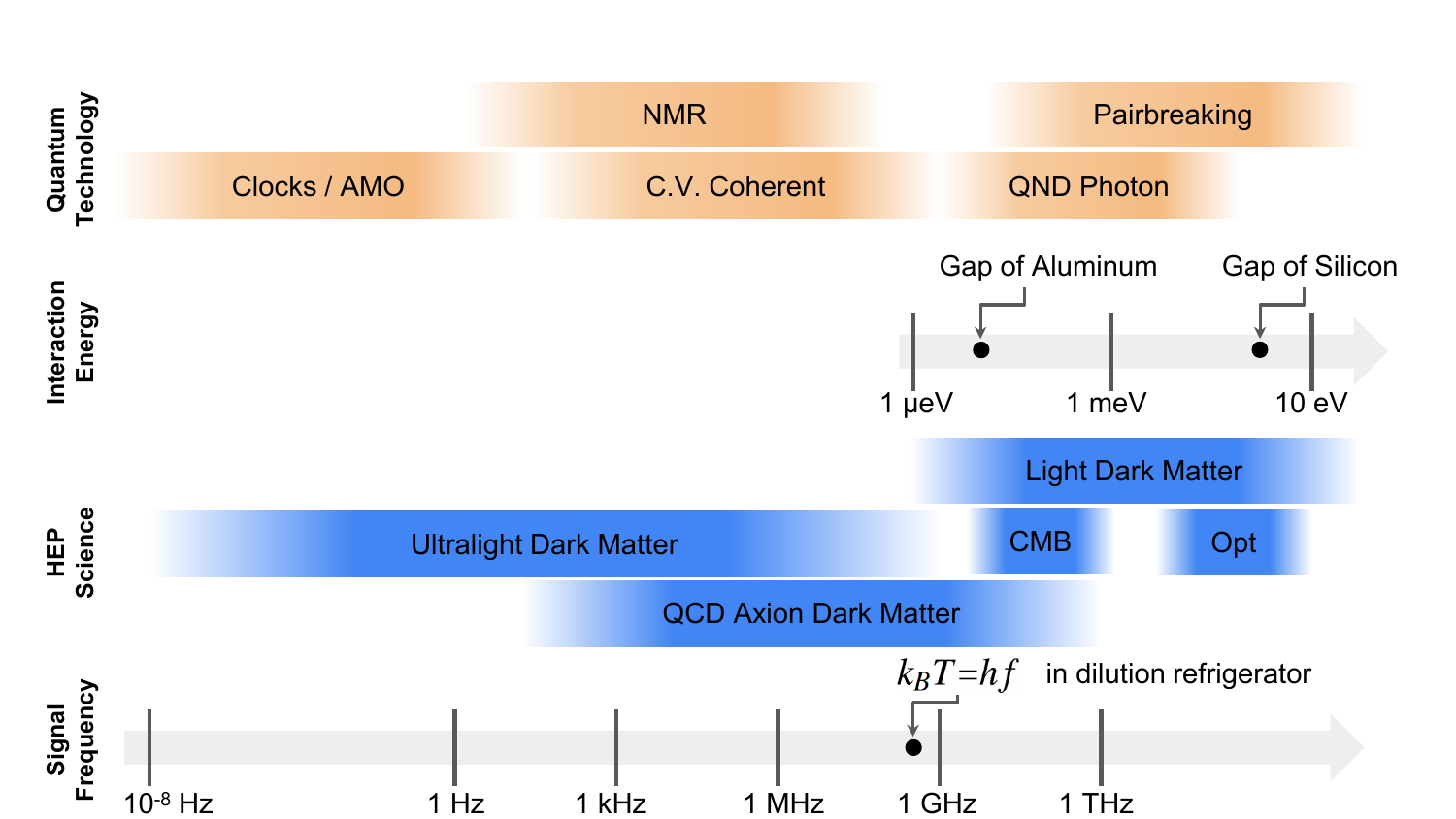}
    }
    \caption{Bottom axis is the frequency of the signal being measured. The middle axis is the interaction energy. Interaction energy and frequency are aligned such that $hf=E$ (although the detected energy $E$ may be from a scattering event and may not include the rest mass of the incident particle). HEP science targets are in blue. Quantum sensor tech categories (in orange) include Clocks / AMO (clocks, ion traps, neutral atom interferometers), nuclear magnetic resonance (NMR) with spin ensembles, Continuous Variable (C.V.) Coherent (including parametric amplifiers, radio-frequency quantum upconverters), Quantum Non-Demolition (QND) Photon sensors ( including qubits, Rydberg atoms), and Superconducting Pairbreaking detectors (including superconducting transition-edge sensors (TES), microwave kinetic inductance detectors (MKIDs), superconducting nanowire single photon detectors (SNSPDs)).  Also indicated are some experimentally relevant scales that define the range of applicability for each sensing paradigm (semiconducting energy gap of silicon, superconducting energy gap of aluminum, and thermal noise frequency where $k_B T = h f$ in a dilution refrigerator).}
    \label{fig:QuantumSensor}
\end{figure}

\begin{figure}[ht]
    \centerline{
    \includegraphics[width=0.95\columnwidth]{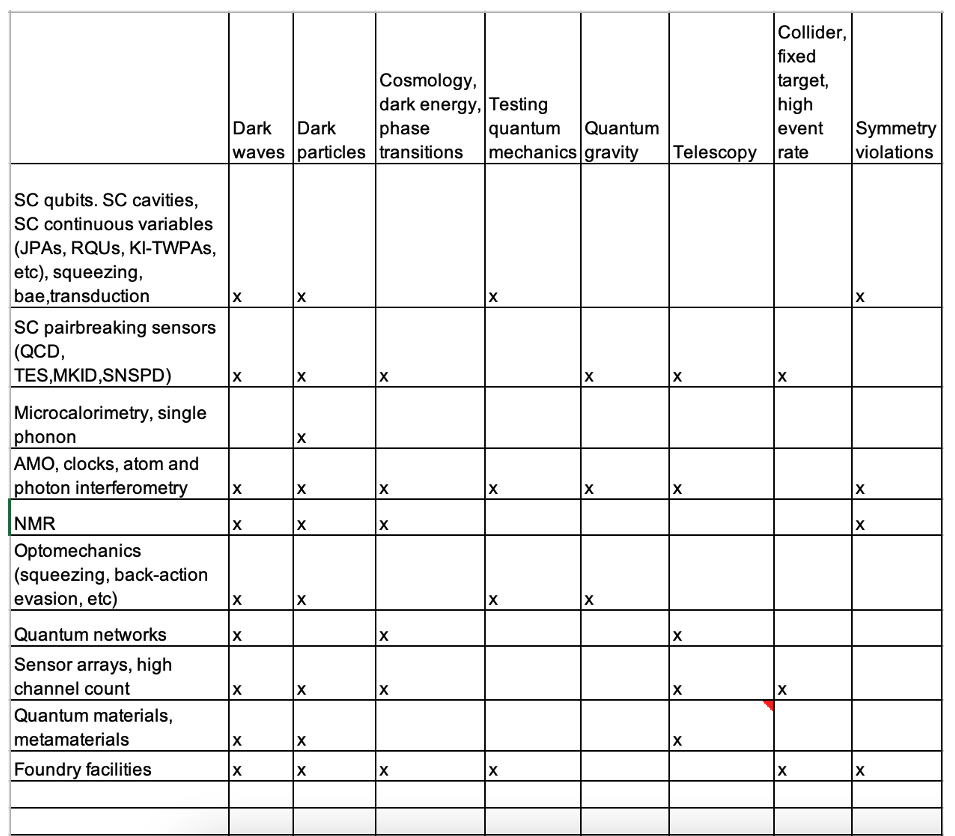}
    }
    \caption{Workshop participants populated this table of HEP science targets versus potentially applicable quantum technologies and explored the use cases at each intersection via key questions listed in appendix~\ref{s:questions}.  }
    \label{fig:table_science_tech}
\end{figure}

\section{Prospective Quantum Sensing Technologies}

\subsection{Atom interferometry}
Atom interferometers exploit spatially delocalized quantum states to look for effects varying between different positions.  In many cases, they additionally leverage quantum superpositions of different internal states to gain sensitivity to effects varying between atomic energy levels.  Atom interferometers offer a pathway to pursuing many of the science drivers described in Sec. \ref{Sec:ScienceDrivers}, including searches for wavelike dark matter (generalized scalars, hidden photons, and axions with various couplings to different standard model particles) in the $<10^{-12}$\;eV range \cite{antypas2022new, kolb2018basic, fleming2019basic}, sub-GeV particle dark matter~\cite{Riedel:2016acj,Du:2022ceh} and dark energy \cite{kolb2018basic,hamilton2015atom,panda2023measuring}; tests for new long-range wave-like interactions between Standard Model particles (fifth forces) \cite{asenbaum2020atom, wacker2010using}; precision tests of the Standard Model \cite{parker2018measurement, morel2020determination}; probes of cosmology via gravitational wave detection in the mid-band frequency range between LISA and LIGO \cite{fleming2019basic, ballmer2022snowmass2021}; and tests of quantum gravity \cite{carney2022snowmass, carney2019tabletop} and of quantum mechanics \cite{abe2021matter,overstreet2022observation,Kaplan2022}.

Two small atom interferometry projects were funded as DOE QuantISED ``Exemplar" experiments.  The Berkeley ALPHA atom interferometer will enable a new generation of precision tests of the Standard Model by making improved measurements of the fine structure constant, building on previous state-of-the-art fine structure constant measurements using atom interferometry \cite{parker2018measurement, morel2020determination}.  The MAGIS-100 100-meter-baseline atom interferometer under construction at Fermi National Accelerator Laboratory will search for wavelike dark matter, test foundational quantum science, and serve as a prototype experiment for atom interferometric mid-band gravitational wave detection \cite{abe2021matter}.  MAGIS-100 builds on advances in large-scale atom interferometry demonstrated at the 10-meter scale \cite{kovachy2015quantum,asenbaum2017phase,asenbaum2020atom,overstreet2022observation}.   While these experiments are examples of shovel-ready atom interferometers, the development and incorporation of new quantum technologies is critical to maximizing HEP science reach.  Quantum control techniques such as dynamical decoupling \cite{Graham2016resonant,zaheer2023quantum} and quantum optimal control \cite{Saywell2018, Chen2023} can enhance sensitivity while suppressing various noise backgrounds.  Moreover, the incorporation of squeezed states can allow atom interferometers to operate at sensitivities beyond the standard quantum limit \cite{greve2022entanglement, malia2022distributed}.  Further development of large-scale atom interferometry, up to the km-scale (MAGIS-100 is a prototype), is also needed—this will require high-flux ultracold atom sources and advanced large momentum transfer atom optics techniques \cite{abe2021matter,badurina2020aion}.  In addition, further advances in ultracoherent optical-lattice atom interferometers can enable new tests of quantum gravity \cite{xu2019probing,panda2023measuring}.

Another important area for research and development is atom interferometric sensor arrays, which have the potential to serve as a powerful tool for rejecting backgrounds and detecting transient signals.  For example, the use of such arrays to suppress backgrounds from Newtonian gravity noise, a dominant source of noise for low-frequency terrestrial gravitational wave detectors, is being investigated \cite{mitchell2022magis, chaibi2016low, badurina2023ultralight}.  The significant global interest and investment in long-baseline atom interferometry—with MAGIS-100 in the US \cite{abe2021matter}, AION in the UK \cite{badurina2020aion}, MIGA in France \cite{canuel2018exploring}, and ZAIGA in China \cite{zhan2020zaiga} all currently under development—presents an opportunity to develop an international network of atom interferometers.  Ultimately, entanglement between different nodes in the network may facilitate dramatic improvements in sensitivity (see Section~\ref{AtomicQN}). Recently, a proof-of-concept demonstration of distributed quantum sensing with an entangled network of atom interferometers was successfully carried out \cite{malia2022distributed}.

Atom interferometers can greatly benefit from the unique capabilities and resources that the HEP community can provide, including: expertise in advanced rf sources for quantum state preparation and control, experience in engineering mid-to-large scale experiments, high-performance computing for large-scale data analysis and simulations (including artificial intelligence expertise), and deep shafts and underground sites at DOE facilities that are necessary for long-baseline atom interferometry.  Furthermore, DOE can (i) pull in expertise from national labs, (ii) keep efforts ongoing for the time scales needed to hunt down systematic effects and solve the integration challenges associated with incorporating QIS techniques with detectors that have search campaigns, (iii) support the collaborations needed for larger-scale projects, and (iv) provide access to needed facilities.

Atom interferometers today are small scale ($<$\$20M), but various science targets may require scaling up.  For example, a full scale ($\sim$km baseline) atom interferometer for mid-band gravitational wave detection would be a larger project.  It is notable that a long-baseline facility designed today could promote research for a future fully-sensitive detector, likely needing only implementation upgrades rather than facility upgrades.

\subsection{Atomic, nuclear, and molecular clocks and optical cavities}
Optical clock precision has improved by more than three orders of magnitude in the past fifteen years, surpassing the microwave cesium clock presently used to define one second by a factor of 100 \cite{Ludlow2015}. The present frequency precision of 1 part in $10^{18}$ corresponds to 1 second over 30 billion years \cite{PhysRevLett.123.033201}.  Dramatic gains in precision put clock operation in uncharted territory to test postulates of modern physics and search for physics beyond the standard model \cite{Safronova2018}. Clocks are already being used for tests of the constancy of the fundamental constants and local position invariance \cite{PhysRevLett.126.011102}, dark matter searches \cite{PhysRevLett.125.201302}, tests of the Lorentz invariance \cite{PhysRevLett.126.011102}, and tests of general relativity \cite{2020NaPho}. Future clock and clock networks development will provide further orders of magnitude of improvement for these experiments \cite{buchmueller2022snowmass,antypas2022new}.

Atomic, molecular, and nuclear clocks are particularly sensitive to some dark matter candidates, uniquely so for ultralight scalars for mass ranges under $10^{-15}$~eV \cite{Safronova2018,antypas2022new}. Ultralight scalars source variation of fundamental constants leading to corresponding variation in atomic, molecular, and nuclear spectra and, therefore, clock transition frequencies. As a result, ultralight dark matter can be detected by measuring the time dependence of frequency ratios of two clocks, a clock and an optical cavity, or two optical cavities. These are direct detection broadband experiments with verifiable signals.

\begin{figure}[ht]
    \centerline{
    \includegraphics[width=0.95\columnwidth]{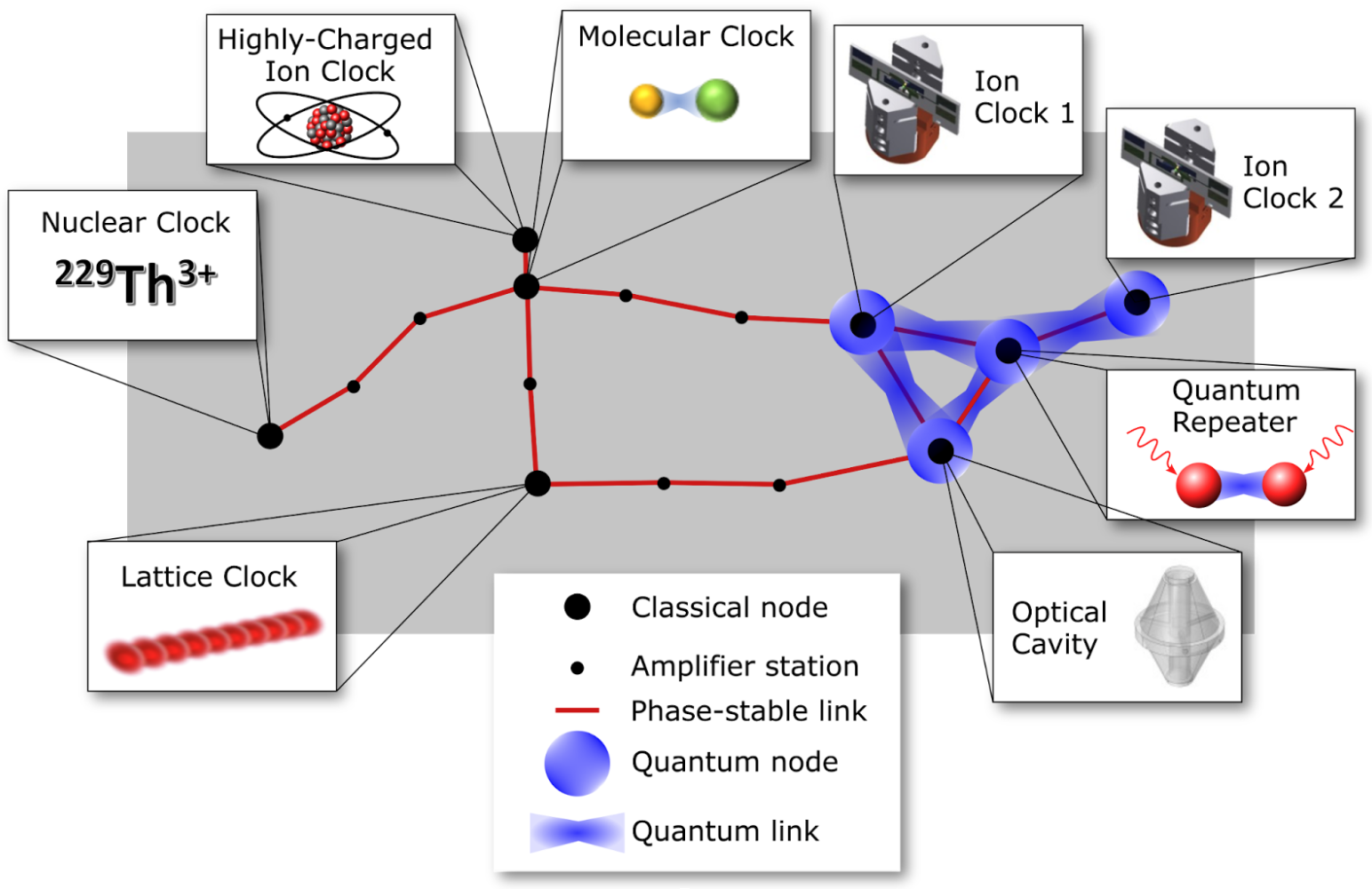}
    }
    \caption{Conceptual picture of a clock network including both classical and quantum nodes and a wide variety of clock species to target various applications in fundamental physics.}
    \label{fig:amo_network}
\end{figure}

Several potential pathways exist for drastically  improving the ability of clocks to detect dark matter and other new physics include:  developing new clocks with much larger sensitivity factors to  dark matter such as highly charged ions \cite{RevModPhys.90.045005} and especially nuclear clocks \cite{nuclear}; development of large and more integrated clock networks \cite{QSNET}; making clocks more portable \cite{2020NaPho}; improving local oscillator technology as it limits coherent integration times; and performing measurements beyond the standard quantum limit. Pushing beyond the SQL can be achieved by using entangled states, such as spin-squeezed states. A conceptual picture of a clock network including both classical and quantum nodes and a variety clock species to target various applications in fundamental physics is given in Fig.~\ref{fig:amo_network}.

Proposed nuclear clocks based on thorium-229 have much higher sensitivities to the variation of the fine structure constant, up to 4 orders of magnitude relative to any other clocks \cite{nuclear}. Nuclear clocks are also highly sensitive to the hadronic sector and could offer improvements in sensing of DM couplings by 5--6 orders of magnitude. 

Deployment of high-precision clocks in space could open the door to new applications, including precision tests of gravity and relativity \cite{FOCOS}, searches for a dark-matter halo bound to the Sun \cite{2023NatAs}, and gravitational wave detection in wavelength ranges inaccessible on Earth \cite{Kolkowitz_2016,Fedderke_2022}. Space-based optical lattice atomic clocks include the possibility of a tunable, narrowband GW detector that could lock onto and track specific GW signals, complementing other experiments (e.g. LISA and LIGO) \cite{Kolkowitz_2016,Fedderke_2022}.
The recent NASA Biological and Physical Sciences decadal survey~\cite{NAP26750} recommended a multiagency  “Probing the Fabric of Space Time (PFaST)” initiative involving deploying high-precision clocks in space. This decadal survey also recommends that “NASA should substantially increase resources dedicated to producing and understanding the answers to the key scientific questions detailed in this report. This investment recognizes the potential for significant societal impacts utilizing the space environment for the biological and physical sciences portfolio in the coming decade, aimed at identifying new principles of physics that can only be discovered through experiments in space, including those governing particle physics, general relativity, and quantum mechanics.”
This recent development provides novel possibilities of multiagency collaboration on HEP applications of quantum clocks. 

The unique capabilities and resources that DOE can provide to advance quantum clock and optical cavities applications for HEP science targets  are similar to those listed when discussing atom interferometers in the previous section, including providing both experimental and theoretical expertise from national labs, solving the  challenges of developing and operating dedicated high-sensitivity clock detectors and networks for HEP science targets, supporting the collaborations needed for larger-scale projects across physics and engineering disciplines, and providing access to needed facilities including high-performance computing and data centers.

\subsection{SNSPDs}
Superconducting nanowire single photon detectors (SNSPDs) are the most advanced technology for ultra-low-noise time-resolved single photon counting~\cite{you20,Wollman21}, and are rapidly advancing in terms of performance. They have demonstrated single-photon counting from EUV -- 29 $\mu$m~\cite{Fuchs22,Taylor:2023cqx}, and array scaling to formats as large as 400,000 pixels~\cite{Oripov:2023yod}. SNSPDs have demonstrated dark counts below $10^{-5}$ counts per second in an 0.16 mm$^2$ pixel~\cite{PhysRevLett.128.231802}, system detection efficiency of 98\% at 1550 nm~\cite{Reddy:20}, timing resolution below 3 ps FWHM~\cite{Korzh20}, maximum count rates above $10^9$ counts per second~\cite{Craiciu:23}. However, further work is required to make individual detectors which meet these metrics simultaneously. SNSPDs are a rapidly evolving technology, with strong potential for further reductions in energy threshold and dark count rate, and scaling toward megapixel arrays and cm$^2$ active areas. SNSPDs have broad applicability to a wide variety of challenges in HEP which require low-noise sensors with low energy thresholds for single excitations, such as searches for particle-like and wave-like dark matter at far-infrared wavelengths. 

\begin{figure}[ht]
    \centerline{
    \includegraphics[width=0.95\columnwidth]{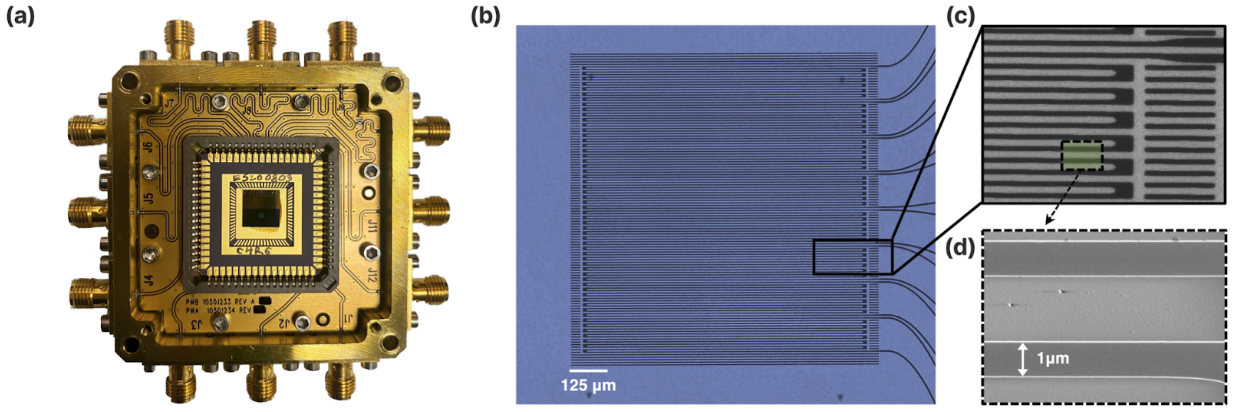}
    }
    \caption{Superconducting nanowire single photon detectors.}
    \label{fig:snspd_images}
\end{figure}

A range of initial science efforts that use SNSPDs to search for dark matter are either underway or planned~\cite{BREAD:2021tpx, Chiles:2021gxk, Hochberg:2021yud}.  The ultralow dark count rate is one of the key features that makes these detectors so attractive. The suitability of these detectors for those efforts suggests the possibility of future application to related challenging detector problems in HEP such as neutron or neutrino detection.

Before SNSPDs can be truly impactful in this domain, however, effort is needed across a number of axes: Firstly, the ultimate source of the residual dark-count rate (however low) must be identified and understood.  At present, dark counts are hypothesized to be due to residual radioactivity in the package and/or cosmic ray events.  Secondly, the detector areas must be scaled up significantly (for some proposed experiments, to m$^2$ areas). Such an effort would require investment in both fundamental device-physics research, and in better understanding of the engineering obstacles that currently face high yield of large area devices.

A major secondary benefit of SNSPDs is their potential use in large-scale detection projects involving high-energy particle tracking in HEP detectors.   SNSPDs have a number of characteristics that would make them extremely attractive as particle-tracking detectors: principally their potential for radiation resistance, their small dimensions, and their excellent timing resolution.  Furthermore, they can be incorporated with electronics that could be used as in-detector event detectors. They are also relatively simple to fabricate, and can be placed onto a number of different kinds of substrates. 

SNSPDs provide a unique opportunity to significantly enhance the detector capability of proposed future collider experiments including the FCC and the Muon Collider. These machines are expected to have orders of magnitude more simultaneous collisions or beam induced backgrounds than current machines. For example, for the FCC-hh, the estimated amount of pileup — the number of simultaneous occurring collision at the interaction point — is  expected to reach 1000, a factor of 20 larger than the conditions currently at the LHC. They will require detectors with a timing precision of 1~ps to assign particles to the correct interaction vertex, reduce beam induced backgrounds, and reconstruct the physics objects on an event by event basis, all necessary to achieve their ultimate science reach.  SNSPDs have demonstrated 2.7~ps time jitter for a single photon and promises to provide next-generation collider experiments with the precision timing requirements needed as a single sensitive element. A strong R\&D program to experimentally demonstrate and optimize the ultimate time resolution performance of SNSPDs to charged particles, expected to deposit larger energies compared to single photons, will pave the road towards understanding the full extent of the physics case for such quantum sensors. 

Additionally, SNSPD are expected to be radiation hard up to the fluences expected for future collider experiments. 
Although published studies in this area are lacking, there is every reason to believe these detectors will be highly radiation resistant.  Two key sources of radiation damage in devices are effects on doping concentrations and displacement defects induced in thin functional oxides (such as the gate oxide of a transistor).  The simplicity and refractory nature of the material mean there are no semiconductors present, thus doping alteration during radiation exposure should not be a concern, and the film is homogeneous, thus displacement damage concern is also minimal.  However, experimental verification is necessary to prove that the sensor’s performance metrics are maintained after radiation expose and this should be an active area of research.

Lastly, implementation of large SNSPD arrays into HEP experiments will require research and innovation into the readout and operation of these large-format SNSPD arrays, especially for low-threshold SNSPDs.

\subsection{Superconducting qubits}

At frequencies where thermal noise is small (corresponding to  $k_B T < h f$, which occurs at frequencies above $f \sim 300$\,MHz in a dilution refrigerator), detecting individual signal photons is an important goal for HEP science. At lower frequencies, the sensing and maintenance of photon-number states (Fock states) can be difficult. Superconducting qubits are an important sensor modality for this application.

Superconducting (SC) qubits including the transmon can be viewed as engineered artificial atoms where the nonlinearity of the 1/R Coulomb potential of real atoms is replaced by the intrinsic nonlinear inductance of the Josephson junction.  LC electronic circuits incorporating Josephson junctions therefore exhibit quantum two-level system behavior which allows these devices to be used in all of the traditional sensing methodologies of atomic physics.  These include direct absorption of photons and also measurements of frequency shifts in atomic clocks due to interactions of the ``atom" with background fields, similarly to how precision electroweak physics infers the presence of Higgs field via precision measurements of the masses of the W and Z bosons. 

\begin{figure}[ht]
    \centerline{
    \includegraphics[width=0.95\columnwidth]{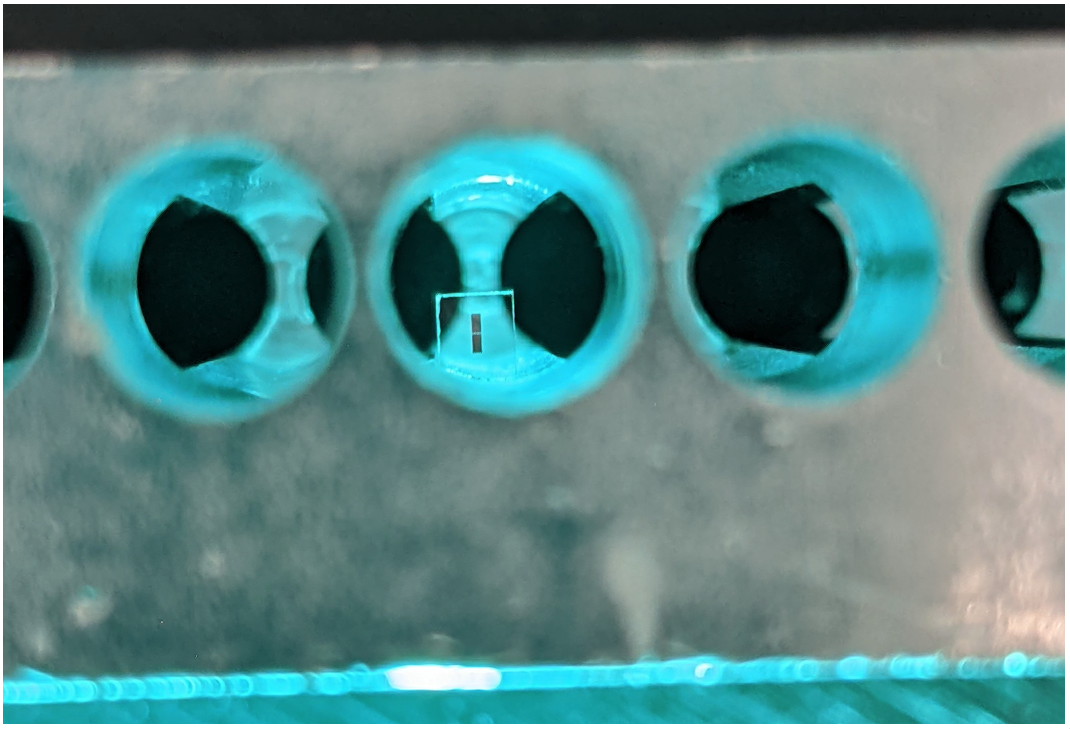}
    }
    \caption{A superconducting microwave ``panflute" cavity with sub-wavelength size holes revealing the superconducting transmon qubit sensor installed within.   The qubit can perform quantum non-demolition measurements on single photons deposited into such cavities by dark matter wave interactions, thus providing the ability detect these ultraweak external forces with sensitivity orders of magnitude better than the standard quantum limit~\cite{Dixit:2020ymh}.  These highly nonlinear Josephson devices can also be used to prepare non-classical probe states such as Fock states, Schrodinger's cat states, and GKP states to boost the dark matter signal rate~\cite{Agrawal:2023umy}.  (Photo credit: Akash Dixit)}
    \label{fig:qubit_cavity}
\end{figure}

Artificial atoms are more versatile than real atoms in that both the transition energy or frequency, and the electric polarizability or antenna size of the atom are now parameters which can be adjusted to match the measurement application -- one no longer has to rely on the limited palette that nature has provided.  For example, while the largest Rydberg atoms have micron-size electron orbitals which couple well to optical photons, the physical metallic antennae of the SC qubit can be much longer, typically mm-scale to provide strong coupling to cm-wavelength photons in microwave circuits.  However it should be noted that mature SC qubit technology is for now only available in the few to 10's of GHz frequency range for which it has been developed by the quantum computing field, and going to higher frequency sensing will require the development of higher~Tc qubits, the use of real Rydberg atoms, or a transition to the pair-breaking sensor technology described below.

A recent QuantISED-funded collaboration of the HEP and QIS communities demonstrated the use of SC qubits to detect single microwave signal photons with the lowest dark rate achieved to date.  In this pathfinder experiment, background dark matter waves scatter on a microwave cavity which serves as the quantum random-access memory (QRAM) for a quantum computer.  The signal is single photons mysteriously appearing in the QRAM, causing bit errors in the quantum computation.  The electric field of the single photon is sufficient to drive a coupled qubit into nonlinear operation, thus inducing a discrete shift in its resonant frequency.  High readout fidelity can be achieved by repeatedly measuring the qubit frequency shift without absorbing or destroying the signal photon. This quantum non-demolition technique evades back-action noise and like other photon-counting detectors, has no fundamental quantum limitation on its noise floor.  The prototype device achieved a factor of 40 suppression of readout noise below the standard quantum limit, and its performance is believed to be limited only by leakage of physical background photons through cracks in the device~\cite{Dixit:2020ymh}. 
Another proposal has been made to skip the microwave cavity and instead use the qubit antennae to directly detect dark photons via the mixing of the dark electric field with the visible electric field~\cite{Chen:2022quj}.  The large noise reduction afforded by these qubit-based single microwave photon counting detectors can speed up resonant dark matter experiments by factors $>1000$ -- a critical speed-up as such experiments are otherwise projected to take $>1000$~years integration time using conventional sensors.

As highly nonlinear devices, superconducting qubits can also be used in conjunction with cavities to produce highly non-classical states of light which are extremely sensitive to environmental perturbations including that from the putative background dark matter.  Due to stimulated emission, a state of Fock number $N$ essentially sees the rate of all interaction processes increased by a factor of $N$.  Combined with high-$Q$ cavity resources, this quantum-enhanced signal can even further improve the sensitivity of wave dark matter searches.  Another QuantISED effort recently demonstrated a factor $>2$ signal enhancement from preparing the cavity in Fock number state~\cite{Agrawal:2023umy}.

While current prototype experiments operate the qubits in situ in the same cavity which collects the dark matter signal from e.g. dark photons, future experiments searching for axions will require the qubit to be placed in a remote location, well outside of the high magnetic field needed for the axion scattering target.  Ongoing efforts target the high efficiency transduction and transport of single photon signals into remote receiving cavities where they can be probed by qubits in a clean, field-free region~\cite{balembois2023practical}.

\subsection{Continuous variables quantum sensors and amplifiers}
At frequencies where thermal noise is relatively large (corresponding to  $k_B T > h f$, which occurs at frequencies below $f \sim 300$\,MHz in a dilution refrigerator), the sensing and maintenance of photon-number states (Fock states) can be difficult. At these lower frequencies, coherent continuous-variables quantum sensors and amplifiers are needed to enable measurement at and below the Standard Quantum Limit\cite{brouwer2022proposal}. Even at higher frequencies where $k_B T < h f$, when searching for a spectrally narrow signal with a wider bandwidth detector (as in axion dark matter haloscopes~\cite{Sikivie1983,Du2018,Brubaker2017}), coherent sensors generally outperform photon counting methods when $N_T \times N_c>1,$ where $N_T=[\exp(h f/k_B T)-1]^{-1}$ and $N_c$ is the ratio of receiver to signal bandwidth. Superconducting parametric amplifiers (SPAs) and RF Quantum Upconverters (RQUs) have an important role in achieving HEP science goals in these frequency ranges to enable sensitive axion searches and other tests of fundamental physics. 

Superconducting parametric amplifiers (SPAs) based on Josephson junctions or nonlinear kinetic inductance can perform quantum-limited measurements of microwave signals~\cite{Yurke1989,castellanos2008} and back-action evading measurements enabling sensitivity beyond the standard quantum limit~\cite{malnou2019, backes2021}. Such capabilities enable sensitive axion searches, both with and without cavities. SPAs can also be used to realize on-chip Fourier transform spectrometers~\cite{BasuThakur2023}, quantum frequency converters~\cite{Bergeal2010}, and sources of entangled microwave photon pairs~\cite{Flurin2012}. While SPAs have been advanced significantly for QIS applications at C-band frequencies (4--8 GHz) such as readout of superconducting qubits, development is still needed in two frequency ranges relevant for HEP: low-frequency parametric amplifiers ($\lesssim 400$\,MHz), and millimeter-wave and terahertz parametric amplifiers (30 - 300 GHz). Recent advances in superconducting materials and nanofabrication have opened the possibility of extending the SPA concept to millimeter and submillimeter waves from 30 -- 300\,GHz. Improvements in auxiliary microwave components as well as new SPA device architectures can extend the performance range of these devices to $\lesssim 400$\,MHz for HEP experiments. 

Radio-frequency quantum upconverters (RQUs) coherently upconvert signals at lower frequencies ( $\sim$\,100\,Hz -- 300\,MHz ) into the microwave C-band (4--8 GHz), where mature superconducting quantum microwave technologies are available. They operate by modulating the frequency of a C-band superconducting microwave resonator by the quantum interference of a Josephson junction array coupled to a low-frequency signal \cite{kuenstner2022quantum}. RQUs are dissipationless and reactive, so are not limited by the thermal Johnson noise that affects the sensitivity of dc SQUIDs, and can in principle achieve operation at the SQL over this frequency range. In addition, they can be operated in a phase sensitive mode, and be used in backaction evasion (BAE), sideband cooling, and other quantum protocols to achieve sensitivity better than the SQL. This phase-sensitive operation promises the sensitivity required for electromagnetic axion searches approaching GUT scale axion masses\cite{brouwer2022proposal}. Initial results with these devices are promising, with strong quadrature-sensitive gain\cite{kuenstner2022quantum}, the first step towards the large BAE contrast that will be required for a GUT-scale axion search. RQUs may also find a crucial application for sensitive readout of spin ensemble dynamics, as described in section~\ref{sec:spin}.

\subsection{Superconducting cavities}

Superconducting cavities are playing an increasing role in QIS and as tools for probing fundamental physics.  These ultralow loss cavities provide superb isolation of experimental systems from the environment, thus enabling long interaction times for signals to be received from coherent but ultra-weak sources and to be stored as quantum information.  As various cavity QED techniques are often employed to enhance the various quantum sensing techniques described above, the extremely high quality factor cavity becomes an integral part of the quantum sensor.  R\&D efforts  in superconducting radiofrequency (SRF) cavities and demonstrations of their utilization in new physics searches are spearheaded within the SQMS NQI center. Some of the progress and the science opportunities are summarized in a Snowmass white paper~\cite{Berlin:2022hfx}.
A few highlights include: 
\begin{itemize}
    \item The Dark SRF experiment has probed previously unexplored parameter space for dark photons (without assuming it is dark matter) in its pathfinder run~\cite{Romanenko:2023irv}. The sensitivity is limited by frequency stability and thermal backgrounds and a second run in milli-Kelvin temperatures is planned~\cite{Contreras-Martinez:2023ovl}.
    \item Deep sensitivity to dark photon dark matter with an ultra-high quality cavity was demonstrated at a single frequency in milli-Kelvin temperatures~\cite{Cervantes:2022gtv}.
    \item Superconducting cavities that can withstand Tesla-level magnetic fields for axion searches can be utilized at a host of Axion DM searches. A custom designed Nb$_3$Sn cavity was recently tested showing near-$10^6$ quality factors at 6 Tesla fields~\cite{Posen:2022tbs}. 
    \item High-Q SRF cavities also provide an opportunity to design high efficiency transducers of single photons from microwave to optical frequencies~\cite{Wang:2022kaa}, opening new possible protocols for photon counting in dark matter searches.
    \item A fixed frequency prototype of a multi-mode SRF based axion search~(see~\cite{ Berlin:2019ahk}) is being developed~\cite{Giaccone:2022pke}.
\end{itemize}

\subsection{Qubit-Based Pair-Breaking Detectors}
\subsubsection{Quantum capacitance detectors}
The Quantum Capacitance Detector (QCD) and similar quantum charge sensors~\cite{Brock2021} are implementations of the single Cooper-pair box (or “charge qubit”) which exploit single-electron charge sensitivity to accurately measure the density of quasiparticles generated by external signals (e.g. far-infrared radiation). QCDs have recently been used to count single photons at 1.5~THz (200 µm, 6 meV) with dark count rates of $10^2$ counts per second~\cite{Echternach2018}. They have been scaled to arrays of 441 pixels, and have been integrated with mesh absorbers and microlens arrays for efficient optical coupling at terahertz frequencies~\cite{Echternach2022}.  As true terahertz single photon counters with a world-leading energy threshold, they have the ability to enable a wide class of fundamental physics experiments probing low-energy excitations, such as axion searches and searches of late dark energy. 

\subsubsection{Superconducting Quasiparticle-Amplifying Transmon}

The Superconducting Quasiparticle-Amplifying Transmon (SQUAT) is based on the transmon qubit, which is a weakly charge-sensitive architecture with tunable separation between charge states. The much smaller charge dependence allows for direct-coupling of a readout line to the device, reducing the amount of passive metal in the design. It also allows for a large collecting area for application to phonon sensing or coupling to photons without the need for additional focusing optics. Initial design suggests sensitivity from the eV to meV scale, potentially down to single broken Cooper pairs\cite{fink2023superconducting}, using quasiparticle trapping similar to the TES-based QETs described elsewhere in this report. Initial tests demonstrating baseline performance should show dark rates comparable to QCDs but with 100x larger collecting area, and with the ability to determine the energy of the scattering event.

\subsection{Kinetic inductance detectors}

Kinetic inductance detectors (KIDs) are pair breaking detectors that exploit the dependence of a superconducting thin-film’s complex impedance on the population of paired (Cooper pairs) and un-paired (quasiparticles) charge carriers. Excitations with energy greater than the Cooper pair binding energy are able to break Cooper pairs and create quasiparticles, modifying the surface conductivity. By lithographically patterning the thin film into a microwave resonator, this modification is sensed using conventional RF techniques to monitor the resonant frequency and quality factor of the resonator. KIDs are attractive as they are naturally frequency-domain multiplexed, where thousands of sensors can be read out on a single microwave line. This feature makes KIDs an important technology for large-scale phonon sensors in future low-mass dark matter experiments where large channel counts could provide improved event reconstruction and background rejection. Similar motivations drive KIDs development for future mm/sub-mm cosmic microwave background and mm-wave Line Intensity Mapping experiments. As a non-linear superconducting device operating at RF frequencies, KIDs have a direct overlap with other quantum technologies including superconducting qubits and continuous variable quantum sensors and amplifiers.

\subsection{Transition edge sensors}
Transition Edge Sensors (TES) utilize small volume superconducting films that are purposefully stabilized within their superconducting transition through electronic feedback as a thermal sensing element\cite{irwin1995application}. Small changes in total electronic energy in this small superconducting volume produce large swings in the resistance of the film which it then read out with SQUIDs (Superconducting Quantum Interference Devices). They have already been shown to have single photon sensitivity for wavelengths $< 10 \ \mu$m  \cite{TES:Fink:2020:WTES}, and and broad band single pixel dark count rates have been measured to be O($10^{-3}$)~Hz/pixel \cite{TES:Shah:2022:ALPS2}.

TES are currently used in HEP for CMB measurements, as well as for dark matter experiments such as TESSERACT and SuperCDMS. To reach their ultimate potential (e.g. as detectors searching for interactions with dark matter in the mass range from 100\,meV to 10\,GeV) requires:
\begin{enumerate}
\item understanding and minimizing both material and external sources of low energy dark counts
\item suppressing environmental parasitic heating that is responsible for limiting their measured single photon sensitivity at $10 \mu$m, $\times20$ higher than their potential theoretical limit of $200 \mu$m (1.5\,THz, 6\,meV)
\item developing materials to consistently and uniformly realize lower T$_{\mathrm{c}}$ superconducting films
\end{enumerate}

\subsection{Spin sensors and NMR} \label{sec:spin}

Spin sensing provides a unique probe of new physics via its effects on spin and electromagnetic dipole moments. For example, nuclear spin ensembles and nuclear magnetic resonance (NMR) are used to search for the defining electric dipole moment (EDM) interaction of the QCD axion~\cite{Budker2014}. Detection of this interaction is necessary to prove that the axion solves the strong-CP problem~\cite{Kim2010a}.
The axion-fermion gradient coupling creates an oscillating torque on the spins of nuclei and electrons. Several experiments make use of ensembles of nuclear- or electron-spin polarized atoms along with precision magnetic resonance detection to search for these interactions~\cite{Aybas2021a,Wu2019a,Garcon2019b}.

Nuclear spin ensembles in non-centrosymmetric solids can search for the defining EDM interaction of QCD axion dark matter over a uniquely broad six-decade mass range, corresponding to frequencies between hundreds of Hz and hundreds of MHz~\cite{Aybas2021a}. The combination of macroscopic ensemble sizes and long coherence times makes this a particularly promising approach~\cite{DeMille2017}. In order to reach QCD axion sensitivity such searches need to solve several technical challenges, which include achieving near-unity spin polarization (which corresponds to purifying the quantum state of the spin ensemble), achieving long spin coherence times, and implementing detection of spin dynamics that is limited by the quantum spin projection noise of the macroscopic spin ensemble, rather than classical thermal or detector noise~\cite{Aybas2021b}. Superconducting quantum sensors and amplifiers, such as the RF Quantum Upconverter (RQU), can play an especially important role in achieving such sensitive detection~\cite{kuenstner2022quantum}. Indeed, coupling such sensors to nuclear spin ensembles may enable quantum engineering of entangled spin states, such as spin squeezing, which can extend the science reach of new fundamental physics searches based on spin ensembles. 

For ensembles of electron spins, microwave cavities or resonators can be used to enhance the spin relaxation rate and to increase the effective volume or antenna-size of the spin. Electron spin excitations and subsequent relaxations can then be sensed by superconducting qubit electronics coupled to the resonator.  For example, single quantum excitations of the collective Kittel magnon mode of a YIG sphere have been detected via the resulting quantized AC Stark shift of a transmon qubit coupled to the hybridized magnon-cavity mode~\cite{TABUCHI2016729}. Single spin excitations of erbium ions have also been detected by absorbing the decay photons using a superconducting qubit as an absorptive single microwave photon detector~\cite{Wang_2023}.

Fundamental science with spin sensors can also reap important benefits from advances in quantum materials. Spin ensemble-based searches for new physics invariably optimize the properties of materials that host the spins. For example, a solid-state search for the EDM interaction of axion dark matter has to identify a crystalline host material that is non-centrosymmetric and contains macroscopic ensembles of heavy atoms with unpaired nuclear spins. These spins need to have a long coherence time, and the material should make it possible to obtain large (ideally near 100\%) spin polarization~\cite{Sushkov2023b}. Isotopic purification can be a straightforward way to improve sensitivity. An experiment searching for the axion gradient interaction with nuclear spin could use a liquid host material, which enables much longer coherence times~\cite{Wu2019a,Garcon2019b}.

Other techniques such as local magnetic field sensing using nitrogen-vacancy or other defect centers, or local spin sensing via the magneto-optical Kerr effect may also be used to detect local spin excitations such as magnon waves.  The defect centers and magnetic two-level system contaminants in materials may also themselves be used as a tracer of energetic dark matter scattering events whose tracks may permanently modify the crystal structure causing discrete shifts in the resonant frequencies of the quantum oscillators~\cite{Rajendran2017,Thorbeck2023}.  Engineering of quantum materials may also be of use in creating bulk targets with tunable spin quasiparticle band gaps to optimize the dark matter scattering kinematics.

A dynamic R\&D program that explores the synergies between quantum engineering of macroscopic spin ensembles, sensitive readout based on quantum electromagnetic sensors, and quantum materials can enable the next generation of HEP searches for new physics~\cite{Sushkov2023c}. 

\subsection{Superfluid helium sensors}
Superfluid helium-4 is a remarkable state of matter: a stable quantum condensate, available in liter quantities, which demonstrates frictionless motion, quantum coherence over macroscopic distances, and quantum interference.  It is available in large  quantities, chemically perfectly pure, easily isotopically ultra-pure, and is routable in both macro and micro-fluidic circuits.  Some of these properties provide important advantages over other quantum condensates such as atomic Bose-Einstein condensates. It is now becoming clear that superfluid helium could be very useful for ultra-sensitive experiments which address fundamental questions spanning geodesy, astrophysics, general relativity, and particle physics.  Superfluid acoustic resonators with very large quality factors (Q~$\approx 10^8$), coupled to very low dissipation superconducting resonators have been demonstrated~\cite{de2017ultra}, with quality factors of $10^{11}$ possible for temperatures $<10$~mK~\cite{de2014superfluid}.  Estimates suggest this system has the potential, limited by thermal noise at milli-kelvin temperatures, to sense the gravitational waves expected from the nearest pulsars~\cite{singh2017detecting} with initial experiments under development~\cite{vadakkumbatt2021prototype}.  Superfluid optomechanics has also been demonstrated over a very large range of sample size, from cm scale to micron size, spanning  $10^3$ to nearly  $10^9$~Hz resonances~\cite{de2014superfluid, harris2016laser, shkarin2019quantum}.  It has now been shown that superfluid acoustic resonators can provide new bounds on the phase-space for dark matter candidates~\cite{schutz2016detectability, manley2020searching},  with initial experiments very recently reported~\cite{hirschel2023helios}.  

Beyond the very sensitive detection possible with extremely low loss acoustic resonators is the opportunity to utilize the quantum phase coherence to form ultra-sensitive matter-wave interference devices.  It has been demonstrated that superfluid helium interferometers with sensing loops approaching 1 meter in length can sense the earth's rotation~\cite{schwab1997detection, bruckner2003large}.  Much more sensitive quantum interference gyroscopes appear possible with the realization of a Josephson junction using new 2D materials, which may put the sensing of the fluctuations in the rotation of the earth's rotation and general relativistic effects such as the Lense-Thirring effect within reach~\cite{packard1992principles}. These devices may find use in precision pointing of telescopes~\cite{phillips2004metrology}. It is also possible that superfluid interferometers may place bounds on dark matter candidates in ways similar to atomic interferometers~\cite{du2022atom}.   Finally, with a superfluid Josephson junction comes the possibility for engineered two-level systems, a superfluid qubit.  This has obvious possible applications to quantum information, but more broadly, as a mechanism for the detection of single acoustic quanta in the superfluid, and the engineering of quantum fluid circuits, similar to what has been demonstrated in superconductivity.  All of these possibilities are waiting to be explored once a proper Josephson junction structure has been demonstrated, an area of current research focus.

\subsection{Optomechanics}
These forms of sensors are fundamentally based on optical interferometry \cite{aspelmeyer2014cavity}. The “mechanics” aspect of these sensors is to use light to probe the position or momentum of a mechanical body that is sensitive to new physics. New physics mechanisms can include: energy deposition from particles, including dark matter \cite{carney2021mechanical,moore2021searching}; coherent transduction from wavelike dark matter or new ``fifth forces'' from controlled sources \cite{manley2020searching,carney2021ultralight,manley2021searching,abbott2022constraints,antypas2022new}; recoil from decay of unstable particles embedded in the mechanical system, for example radioactive isotopes in search of sterile neutrinos \cite{carney2023searches}; gravitational acceleration of the suspended masses from stochastic cosmological or quantum gravitational backgrounds \cite{hogan2008measurement,chou2017holometer,scientific2019search,verlinde2021observational}. 

The unique advantage of optomechanics is probing large, macroscopic masses at their quantum limits of sensing their motion. Optomechanics can be used to deeply search for forces on the test masses delivered from gravitational or BSM couplings into heavy masses. Heavier mass sensors have the advantages of large, kg-scale, detection payloads. Lighter masses have advantages of ultra-low energy thresholds, multiplexed fabrication like MEMs devices, and multiplexed readout if developed in parallel with continuous-variable QIS that also utilizes multiplexing. Multiplexed sensors and readouts allow for the detection of coherent signals such as long-wavelength excitations and particle tracks while rejecting backgrounds.

Optomechanics is one of the first areas to strongly benefit from non-classical quantum state preparation, through squeezed light, as particularly demonstrated by LIGO \cite{mccullerPRD21LIGOQuantum, tsePRL19QuantumEnhancedAdvanced, yuN20QuantumCorrelations}. Further developments in the area will include developing techniques to multiplex sensors \cite{carney2020proposal,carney2021mechanical}, apply squeezed light to sensor arrays with low optical losses required to benefit from quantum enhancements \cite{brady2023entanglement}, and finally to develop back-action evading techniques to probe masses deeply below quantum measurement limits \cite{chen2011qnd,ghosh2020backaction}. In addition to HEP BSM goals, macroscopic objects probed at their quantum limits generate and utilize non-classical states of large-scale systems. Further development of the assembly and preservation of such large scale quantum states enables new tests of quantum mechanics and of gravitational interactions of quantum mechanical systems \cite{moore2021searching,carney2019tabletop}.

\subsection{Quantum networks and long-distance quantum coherence}
\begin{figure}[ht!]
    \centerline{
    \includegraphics[width=0.98\columnwidth]{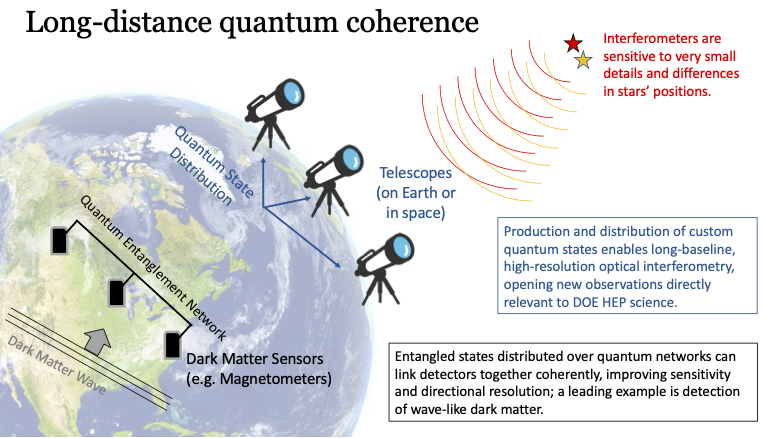}
    }
    \caption{Illustration of extended sensor arrays connected by quantum networks.  Such networked arrays can provide significant advantage in observing an influence which is coherent across long distances, greatly improving both sensitivity and spatial/angular resolution.  The examples shown here are light from distant stars and wave-like dark matter, which directly address HEP science goals; many other applications are also possible.}
    \label{fig:Quantum_Netowrk}
\end{figure}

An important emerging quantum technology is the ability to coherently create, transmit, store and process entangled, non-classical states of multiple qubits separated across macroscopic distances, from tens of meters to hundreds of kilometers.  This capability is generally called “quantum networking” or, more technically, creating and maintaining “long-distance quantum coherence”.
Long-distance quantum coherence will, generally speaking, enable separated arrays of quantum sensors to act as one coherent unit, providing great advantages for (i) detecting weakly interacting radiation with long wavelength, (ii) detecting radiation arriving in single quanta, and (iii) reconstructing the direction of radiation to high precision by using long baselines.  Detection of wave-like dark matter is an example of (i), and precision astrometry, e.g. sky position measurement, of astronomical objects is an example of (ii) and (iii), which are directly relevant to HEP science goals.  Both of these are illustrated in Figure~\ref{fig:Quantum_Netowrk} and discussed in detail below, followed by discussion of specific technology needs and existing capabilities/resources in DOE facilities.

\subsubsection{Optical interferometry and precision astrometry}
The highest angular resolutions achievable in astronomy, either for imaging, e.g. object features, or for astrometry, e.g. object positions, are reached through interferometry of photons from a source.  Traditional methods descend from Michelson stellar interferometry, receiving one photon at a time and interfering it with itself, and Hanbury Brown \& Twiss (HBT) intensity interferometry, interfering a pair of photons together.  Both Michelson and HBT interferometry can be enhanced and expanded through the application of controlled, distributed quantum states, in particular to allow longer baselines and even higher precision in angular resolution. 
Some HEP science goals which would benefit directly from such improvements are discussed in Section~\ref{subsubsec:improved_sky} above, and include $H_{0}$ tension, GR tests, galaxy and structure formation and early-universe phase transitions.

A very basic technology development needed for quantum-assisted interferometry is improved single-photon detectors.  Advantage can be gained from having better time resolution, cost per channel, low maintenance, spatial density, and photodetection efficiency (including sensitivity in the infrared).  Closely related is the desire for a {\em fast spectrograph}, i.e. the ability to disperse a single-mode photon beam by wavelength out onto an array (either 1-D or 2-D) of independent detector elements, each of which is single-photon sensitive with good timing resolution ($<100$~psec). In practice, we aim for a single-photon sensitive device that can both time-stamp and frequency-stamp each detection event, for example Ref.~\cite{Jirsa_arxiv2304}. Building these kinds of detectors has historically been a core HEP capability and systems addressing these requirements have considerable overlap with other quantum sensors discussed in this document, such as SNSPD's deployed in arrays. 

The idea for a quantum-enabled long-baseline telescope system was initially proposed by Gottesman et.al. \cite{Gottesman_2012} (GJC scheme), a proposal based on quantum repeaters. An elementary proof-of-principle can be found in Ref. \cite{brown2023interferometric}. An alternative approach that greatly simplifies the technical requirements while still promising an advantage over classical systems is the SNSV scheme \cite{Stankus_2022}, whose proof-of-principle demonstration was recently implemented \cite{crawford2023quantum}. Both the GJC and SNSV schemes will benefit from exploring spectral binning, for instance, by using fast spectrometers as discussed previously. The detector and spectrometer technology developed within the quantum-assisted telescope lines could also benefit HEP and other quantum applications. Further development is necessary to achieve the goals of quantum-enhanced astrometric measurements, which in turn will benefit the HEP cosmology mission, in particular contributing to solve the Hubble tension, to enable gravitational wave detection at different frequencies than LIGO or NANOGrav~\cite{NANOGrav2023} and to the other areas listed in Section~\ref{subsubsec:improved_sky} and Fig.~\ref{fig:table_science_tech}.  

It should be remembered, though, that in the case of something like a quantum-enhanced optical interferometer it is the entire system, not just the core detector element, that constitutes the ``quantum sensor''; see the original conception of Gottesman, et.al. \cite{GJC_PRL2012} and recent experimental work by Brown, et.al. \cite{brown2023interferometric}.
Accordingly there is a whole set of technology needs for single photon handling, particularly long-distance transport with spatial and polarization mode control.  This would include capabilities such as monitoring and compensating length changes in long fibers, and maintaining sub-nanosecond synchronization across kilometer distances, which are well developed in HEP as part of accelerators; less familiar but also interesting would be developing fast and low-loss fiber switches and solid-state length compensation systems.

\subsubsection{Dark matter detection and quantum networked sensors}\label{AtomicQN}
Deploying a network of (independent) sensors allows researchers to study information not possibly attained by one isolated single detector. It is not surprising that classical networks of atomic sensors are currently being developed to advance the search for fundamental physics. For example, the Global Network of Optical Magnetometers for Exotic (GNOME) physics searches is a collaboration of several research groups with multiple deployed magnetometers around the world with the principal goal of detecting topological defect DM in the form of transient signals from the domain walls of axion-like particles~\cite{GNOME1,GNOME2}. The Networked Quantum Sensors for Fundamental Physics (QSNET) collaboration in the UK is a network of seven atomic and molecular clocks of different species designed to search for deviations in the fine structure constant and the electron-to-proton mass ratio~\cite{QSNET}. In addition, there are two planned networks of atom interferometers for searches of gravitational waves and/or ultra-light DM: the Atom Interferometric Observatory and Network (AION) collaboration is the UK~\cite{badurina2020aion} and the Zhaoshan long-baseline Atom Interferometer Gravitation Antenna (ZAIGA) collaboration in China~\cite{zhan2020zaiga}. 

A network of sensors has a better intrinsic sensitivity than a single detector, and differential measurements can offer common noise subtraction. Distributing entanglement among the different nodes builds on further sensitivity improvement and opens the door for quantum metrology techniques~\cite{QMetrology}. If these quantum networks of atomic sensors are distributed over long baselines, single events produced by long wavelength effects (such as topological defects DM or gravitational waves) can be correlated and positively confirmed. Quantum entangled sensors are fundamentally more sensitive than classically-correlated sensors. Using an entangled network of $M$ independent sensors could offer an advantage in sensitivity of up to $\sqrt{M}$, in addition to a detection bandwidth improvement, when compared to a uncorrelated network of sensors~\cite{QClocksLukin}. This limit in sensitivity is known as the Heisenberg Limit (HL), and it is the lowest bound where particle number detection can be made with minimal (unity) uncertainty. Saturation of such a bound is an on-going effort in the scientific community. Some experiments have recently reported the creation of short-distance entanglement between atomic systems using spin-squeezing~\cite{ClocksNet1,ClocksNet2,malia2022distributed} or using the Duan-Lukin-Cirac-Zoller (DLCZ) protocol~\cite{DLCZ1,DLCZ2}. However, three main research advances are still needed for potential benefit to the HEP community: (i) how entanglement can be distributed directly onto the degrees of freedom used for different atomic sensing protocols (magnetometers, optical clocks, interferometers), (ii) demonstration of sensitivity advantage under realistic deployable environments; and (iii) leveraging such quantum advantage for long-baseline ($\sim$km) networks. This latter point would likely require a first demonstration of a quantum repeater-like configuration connecting distant atomic sensors.    

Some initial funding from the National Quantum Initiative (NQI) and from the DOE-ASCR office is directed towards quantum networking. In addition, there exist quantum network testbeds in a few DOE national laboratories that can be tailored towards entanglement distribution on matter qubits and thus can contribute to HEP quantum sensing needs. A large scale investment could be initiated to create (classical) networks of sensors, focused mainly towards infrastructure and network control development. In parallel, distributed quantum sensing demonstrations exploiting entanglement advantage in small scale experiments should be studied for subsequent quantum enhancement of such long-baseline networks.    

\subsection{Quantum materials}
A large effort has been ongoing (e.g. in DOE-BES) to develop and understand materials where symmetry-based topological concepts give rise to new properties and behaviors. Discoveries in quantum materials include topological insulators and superconductors, massless Dirac fermions, Weyl fermions, the spin Hall effect, spin-orbit torque, chiral magnons, electric-dipole spin resonance, skyrmions, spin-charge conversion, etc. This wide range of properties and rapidly improving fabrication techniques for materials displaying them has great potential for applications in HEP. For example coupling of dark matter models to coherent modes with the correct symmetries and exploiting some of these effects to invent new kinds of sensors. As discussed in section~\ref{sec:spin}, material optimization is also vital for HEP searches based on spin sensors. Collaboration with condensed matter physicists and materials scientists is essential.

\section{Theory}
A robust theory program will be crucial to the HEP quantum sensing strategy.  The role of theory in developing the emerging quantum program has been summarized in a Snowmass report~\cite{Catterall:2022wjq}. Most of the current ideas for utilizing quantum sensors for HEP science targets have arisen from the HEP theory community who have been working closely with experimentalists to study the HEP requirements and the capabilities and limitations of quantum sensing technologies. We refer to the Snowmass report above for many references.  In some cases, an understanding of the new sensing capabilities to probe nature in completely new ways has inspired new theory model-building to expand the range of possibilities of BSM physics.   For this relatively new area of quantum-enabled HEP science, it does not make sense to compartmentalize efforts, but rather encourage active engagement of scientists across all disciplines.  Continued theoretical activities would include:
\begin{itemize}
\item Development of new experimental methods to detect new physics, across technological fields (see section 3 of~\cite{Catterall:2022wjq}).

\item Development of theoretical understanding of science targets for quantum experiments, such as mapping out the space of current constraints on proposed quantum experiments to understand which quantum experiments are well motivated. This activity also includes work on understanding the full theoretical landscape of possibilities from which new physics could emerge. 

\item The development of theoretical frameworks that provide new science targets for quantum experiments, such as new potential signatures of dark matter, dark energy, gravitation and quantum mechanics

\item Development of new algorithmic methods and protocols to improve quantum sensing, including resource distribution across quantum networks, protocols for back-action evasion, photon counting, etc.
\end{itemize}

\begin{figure}[ht]
    \centerline{
    \includegraphics[width=0.5\columnwidth]{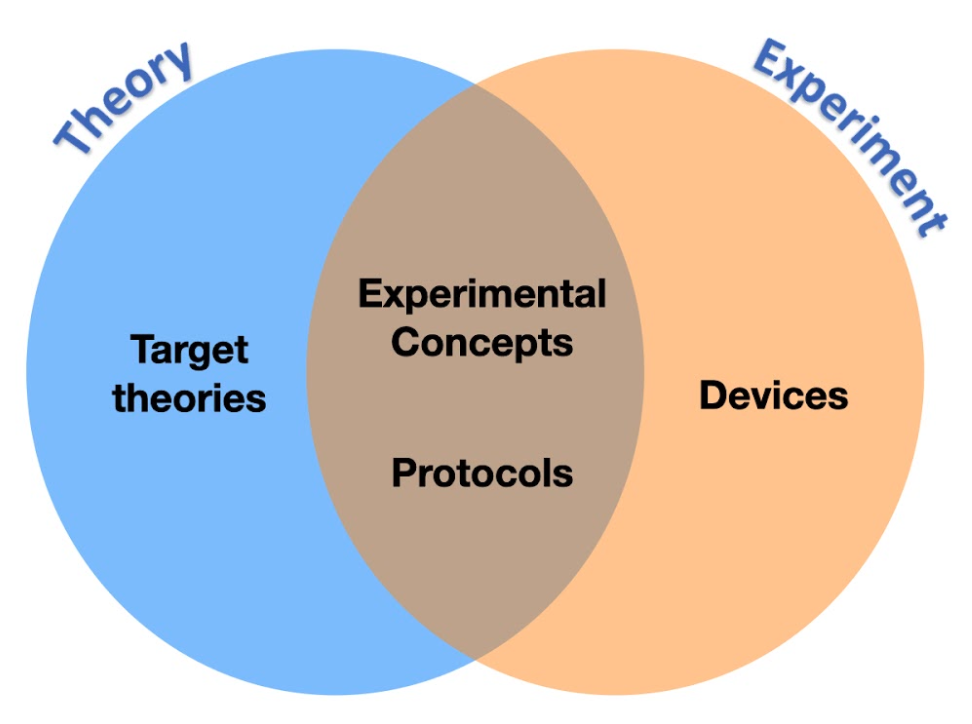}
    }
    \caption{The new quantum frontier requires close collaboration between theory and experimental efforts.}
    \label{fig:theory_expt}
\end{figure}

\section{Common challenges, needs, and remaining gaps in the current R\&D strategy}
\label{s:challenges}
Utilizing new quantum sensing technologies in HEP experimental settings will yield immediate results in the harvesting of low hanging fruit, by searching for new physics in regimes in which nature has never yet been probed.  Indeed, in many cases, the demonstrated successful performance of quantum devices operated in their original applications, such as trapped ion computing platforms, can already be used to set limits on the strengths of possible dark matter interactions with the various sensor materials which would have perturbed such operations.  In other cases, deployment in specialized HEP experiments and optimizing the parameters of the quantum sensor design are necessary in order to maximize sensitivity to the predicted BSM signals while minimizing potential backgrounds which were not of concern in the original applications of the quantum device.

In addition to the challenge of constructing suitable quantum sensors is the challenge of creating experiments that can fully utilize this performance. For photon counters, this means taking advantage of the very low detector dark-count rates by not introducing experimental parasitic dark counts.
For spin sensors, this means suppressing external classical noise so that the spin dynamics readout sensitivity is limited by the intrinsic quantum noise of the spin ensemble. 
For optical interferometry, this requires demonstrating very high levels of contrast. Successful implementation enables protocols that bypass the SQL set by shot noise, even approaching the Heisenberg limit. Another example is low-threshold energy deposition detection, where substrate micro-fractures create background events. Further development of materials processes and background rejection through sensing of multiple signals is needed. Similarly, the effective use of coherent sensors at and below the SQL in the 100\,Hz -- 100\,MHz frequency range requires experiments that carefully control thermal fluctuations by maintaining very low temperatures, and greatly reducing coupling to loss (for example, by the use of very high Q resonators decoupled from loss sources). Successful implementation enables the use of protocols that bypass the SQL set by vacuum noise and by quantum backaction. Lastly, the majority of the envisioned superconducting quantum technologies  implement cryogenic readout including first stage low-noise amplification and cryogenic ``edge processing'' (e.g. multiplexing). In many cases, these ``readout'' technologies (both room temperature and cryogenic) need to be tailored to the quantum sensor in order to realize the quantum gain from the technology.

Another scenario is one in which the quantum sensing techniques, while viable in their original settings, simply are not yet able to achieve the required performance specifications for the HEP application.  Alternatively, there may be other practical impediments to the immediate deployment of new quantum sensing techniques in HEP applications.  In some cases, such as in quantum networking or in the high fidelity operation of quantum computing qubit arrays, the quantum technology is still a work in progress.  In other cases, the required infrastructure to produce the quantum materials or the quantum devices does not yet exist or is not easily accessible or scalable.  Similarly, scalable electronic control and readout technologies will also need to be developed to advance from proof-of-principle quantum sensing demonstrators to the complexity of full-scale HEP experiments.

This section summarizes some common challenges, needs, and gaps in the current Quantum Sensors for HEP effort as identified by the workshop participants.

\subsection{Covering large ranges in signal energy scale}
An intrinsic aspect of sensor technology is that any given sensor will have high quantum efficiency over a limited range of energy or wavelength scales.  The underlying reason is simply wave interference -- the Born approximation indicates that the scattering matrix element is proportional to the spatial Fourier transform of the scattering element, with respect to the momentum transfer.  So the antenna size of the sensor must be matched to the wavelength of the signal in order to obtain high efficiency for detecting the full amplitude of the signal wave.

This limitation then implies that to cover large ranges in signal energy scale, for example the 50+ orders of magnitude in possible dark matter masses, a similarly large range of sensor technologies will need to be studied, developed, and engineered.  Not only does the physical size of the sensor need to be optimized, but the underlying physics of the quantum transitions coupled to the antennae will also need to be matched in energy scale to the HEP process targeted by the experimental application.  The community must develop the necessary collection and coupling strategies to efficiently transfer these small signals to the quantum sensor for measurement (superconducting antennas, anti-reflective coatings, phonon and quasiparticle collection efficiency, spin sensors, etc).

For example, there is a critical need for low noise single photon sensors in the 10-100~GHz range where there a big hole in the frequency coverage for low mass dark matter signals.  While bolometric transition edge sensors operate in this frequency range for large, dc power loads from the cosmic microwave background experimental applications, their noise equivalent power is orders of magnitude too large for the low photon rates expected in axion search experiments.  Meanwhile, the single photon counting techniques employing superconducting qubits will be difficult to scale as the devices will become lossy as frequencies increase towards the $\sim 100$~GHz Josephson plasma frequency in aluminum junctions.  Possible R\&D strategies would be to either make qubits out of higher-$T_{\text C}$ superconductors such as AlN with larger plasma frequencies, or to make low heat capacity TES's or other Cooper pair-breaking sensors out of lower-$T_{\text C}$ superconductors such as AlMn or by reducing the $T_{\text C}$ via the proximity effect.    

There is a similar need for coherent sensing (including continuous variables quantum sensors) operating at or better than the Standard Quantum Limit (SQL) at frequencies from 100\,Hz -- 100\,MHz. Existing sensors operating from 100\,Hz -- 100\,MHz, including dc SQUIDs, are a factor of a few away from the SQL.

Similar considerations apply to the challenge of detecting sub-eV dark matter recoils or absorption events since the electronic band gap in semiconductor materials is eV scale.  While the readout of these particle detectors can be made of sub-meV band gap superconductors, for dark matter scattering events mediated by kinetically mixed dark photons, the bulk target material should be some kind of insulator which would not suppress the electromagnetic response as would happen in the bulk of a superconducting target.  Research is ongoing in studies of both conventional materials containing gapped optical phonon modes as well as engineered quantum materials that have similarly small sub-eV band gaps.  Materials purity will be an important issue as such exotic targets will not enjoy the quality control achieved from the large industrialization of semiconductor production.

Such technology development is critical for filling in under-explored frequency regimes in dark matter searches. No other science areas covered by other technology development sponsors (Astronomy, QIS, remote sensing, communications, etc) have comparable noise requirements in these frequency ranges, making this a necessary area for investment in targeted technology development to enable HEP science. 

\subsection{Operating quantum sensors in high magnetic fields}
Most quantum sensors described in this document have some degree of incompatibility with operation in the high magnetic fields required in HEP experiments such as axion searches or particle tracking in collider detectors.  Indeed, the physical sub-eV energy scales employed in most quantum sensors can be traced to spin interactions such as the Pauli exchange energy or the Zeeman energy which are typically sub-eV in ambient field-free environments.  

In the presence of large background magnetic fields, these quantum sensors will cease to work due to either quantum phase transitions in the underlying materials or due to energy and frequency scales becoming so large that the devices can no longer be easily or economically controlled.  For example, above $H_\mathrm{c1}$ of Type II s-wave superconductors, the Pauli limit is violated as the Zeeman energy exceeds the binding energy of the Cooper pairs.  The antisymmetric spin-singlet Cooper pair state then breaks as the electron spins are forced to align with the external magnetic field.  Sensors made of thin film superconductors with a high upper critical field $H_\mathrm{c2}$, for example SNSPDs made of NbN and other superconductors have been shown to operate with satisfactory performance in parallel fields as high as 6~T~\cite{Polakovic2020,Lawrie2021}.  However most superconducting devices are made of lower $T_{\text C}$ and lower $H_c$ materials, though there has been promising research into Josephson devices utilizing higher-$T_{\text C}$ materials, for example for THz frequency devices~\cite{Borodianskyi2017}.  This triplet superconductivity is no panacea as the breaking of time reversal symmetry removes the protection from impurities guaranteed by Anderson's theorem for s-wave superconductors.  However, breaking of rotational symmetry by the order parameter also produces an anisotropic superconducting gap which may be of use in directional detection of dark matter~\cite{Boyd:2022tcn}. 

Similarly, the operation of sensors based on clock transition frequencies are also greatly perturbed by background magnetic fields.  Atom and ion lattices and interferometers require high isolation from non-uniform external magnetic fields in order to maintain their high timing accuracy.  The hyperfine splitting in color centers such as the nitrogen vacancy center would also increase from GHz microwave frequencies to THz where control electronics and signal sources become significantly more complicated and expensive.  It remains an ongoing challenge to engineer these artificial atoms to provide strong coupling to milli-eV scale excitations.  

Besides R\&D into sensors which are operable in high fields, another avenue of research is to transport the signals out of the high field scattering region into magnetically-shielded, low-field regions where existing sensors can be operated.  The crudest examples of such schemes are ones in which the signal photons slowly leak out and are collected in some remote location with some efficiency which can be degraded by losses in the transport.  In more advanced techniques, the modes in the near and far regions are hybridized to obtain greater quantum control of the states;  these schemes are far more sensitive to transport losses which degrade the mixing of the two modes.  In both cases, techniques from quantum networking technology may become useful, for example by transducing signal photons to other frequencies which are more amenable to transport or measurement~\cite{Wang:2022kaa}.  R\&D in this area would be highly synergistic with quantum networking technology development as both applications require transduction efficiencies far higher than currently available.

\subsection{Materials}
\subsubsection{Quasiparticles}

In superconductors, Bogoliubov quasiparticles are the charge excitations of the Cooper pair condensate, created when Cooper pairs are broken due to thermal fluctuations or due to transient deposits of energy which exceed twice the superconducting gap.  In more general quantum materials including superfluid helium, one may view quasiparticles as the elementary excitations, both charge-like or flux-like, above the ground state of the system.  These quasiparticle states form a heat bath which may cause devices to become lossy.  Alternatively, directly detecting the quasiparticle excitations, often occurring with sub-eV band gap, provides an important sensing channel for reaching low energy thresholds.

A mysterious non-equilibrium population of quasiparticles has been found in all superconducting devices of order one quasiparticle per cubic micron.  Fluctuations in this athermal population creates noise in kinetic inductance detectors due to the corresponding fluctuations in normal and supercurrent components. These quasiparticles also create charge noise in superconducting qubits where scattering of charged quasiparticles on the Cooper pair condensate can excite the Josephson oscillator to higher excited states or extract energy from the oscillator causing reduced lifetimes and coherence times.  When used as clocks, the qubits then suffer from frequency noise just as the kinetic inductance detectors do, and when used as absorptive pair-breaking single quantum detectors, the quasiparticle events give a background dark count rate.  

It is unknown whether the quasiparticle population is of exotic or conventional origin, but recent studies have shown that a combination of cosmic rays, ionizing radiation, and release of mechanical stress in solid state devices create a resolvable rate and spectrum of discrete quasiparticle burst events which may explain the non-equilibrium population~\cite{Mcewen2022,Thorbeck2023,Mannila2022} and its catastrophic effects on quantum computers.  Further efforts on mitigating these conventional backgrounds would result not only in reduced backgrounds for rare event searches at low energy thresholds, but also in more robust operation of solid state quantum devices such as superconductor- or semiconductor-based quantum computers.

The above quasiparticles are often produced by phonon absorption in a superconductor- typically Al. Since the physics signal of interest couples to these phonons (and not the quasiparticles directly), the science cares not only about quasiparticle dynamics, but also about the transduction mechanisms from phonons to quasiparticles. Physical mechanisms include: phonon transport in materials and across interfaces, phonon dispersion relations in materials, phonon lifetimes etc. These are all condensed matter topics with deep expertise in that community. Some connections have been made via QuantISED, but much more collaboration is needed to reach the level of phonon control needed e.g. for kg-yr DM searches. The main gap in this area is materials expertise including: how to make lower noise dielectrics, realizing more uniform and lower $T_{\text C}$ superconducting films, and phonon engineering of different material surfaces for reflection and absorption. Relevant device level calculation and simulation, along with experimental studies are required to advance our understanding and eventual control of these topics. The necessary tools for phonon kinematics simulation and device level calculation is limited to the existing framework of Si and Ge based materials in the current context, therefore investment of significant effort in developing and expanding such tools to cover a broad range of materials and platforms is essential.

\subsubsection{High-$T_{\text C}$ Materials}
High-$T_{\text C}$ superconducting materials such as magnesium diboride (MgB$_2$) are relevant across a variety of different technology areas. Superconductors with a higher gap energy can have lower RF loss at millimeter wave frequencies, allowing for the development of parametric amplifiers and kinetic inductance based qubits at millimeter wave and submillimeter frequencies (30 – 300 GHz). Such materials can also lead to SNSPDs and superconducting electronics with higher operating temperatures (10 – 25 K) which can greatly reduce cryogenic requirements (and thus cost) in large-scale applications.

At present, a few initial results suggest single-photon detection should be possible in high-$T_{\text C}$ materials such as MgB$_2$~\cite{Cherednichenko_2021}  and BSSCO~\cite{Charaev2023}.  
In addition, recent efforts have shown that helium-ion microscopes can be adapted to perform lithography and thus form wires and even Josephson junctions\cite{Kasaei2018} using these materials. 
While initial successes are encouraging, there are key challenges remaining for their wider adoption including the realization of ultra-thin films necessary to fabricate such sensors and ensuring the quality and uniformity of the films. 

\subsubsection{Phonons and phonon engineering}
Since quantum sensors are ideal for harnessing sub-eV energy from DM interaction with target materials, systems capable of producing sub-eV optical phonon excitations (e.g. diamond, SiC, sapphire, silicon) have a great potential to sense sub-MeV dark matter signals when combined with a phonon or a quasiparticle sensor. Furthermore, in order to utilize the full potential of qubits and superconducting sensors as detectors, a thorough understanding of generation of quasiparticles and phonon kinematics in the qubit itself is necessary for application specific materials and platforms along with the device simulation. Particle simulation tools like Geant4 \cite {AGOSTINELLI2003250, ALLISON2016186} along with the Geant4-based toolkit G4CMP \cite{kelsey2023g4cmp, osti_1862281} for phonon kinematics simulation can be implemented for modelling the detector behavior to understand physical mechanisms like phonon transport in materials and across interfaces, phonon
dispersion relations in materials, phonon lifetimes etc. However, the existing modeling capability of G4CMP is limited to Si and Ge and much work is needed to expand it to other materials (e.g. 3C-SiC, diamond in progress) and platforms. Therefore, parallel efforts in developing such toolkits are necessary along with detector development, in order to realize the full potential of novel materials and quantum sensors. 

In addition, quasiparticle poisoning due to pair-breaking phonons is a problem in many quantum devices and sensors, especially in cases where substrate phonons are an unwanted background. This includes many types of photon detectors applied to HEP applications such as TES and MKID sensors, as well as all conventional applications of superconducting qubits. Work is needed to better understand ways to reduce phonon coupling to a variety of superconducting films across a number of substrates. Much of this work can be enabled by the tools mentioned above, but new tools to better model, design, and test phononic bandgaps and phonon filters is needed to further enable meV-scale sensing technologies and produce more radiation robust photon sensors.

\subsubsection{Microfractures}
Searches for interactions of GeV dark matter with calorimeters have been limited by an unexplained source of low energy ($< 1$~keV) background events [CRESST \cite{TES:CRESST:2022:LEE}, CDMS CPD \cite{Microfractures:CDMS:2021:CPDv1}, EDELWEISS \cite{Microfractures:EDELWEISS:2016}]. Most interestingly these events have been seen to:
\begin{enumerate}
\item have significant rate variation with time since cooldown
\item are non-ionizing for even the highest energy depositions
\item have rates that scale with sensor volume or detector surface area rather than detector volume. 
\end{enumerate}
A hypothetical source that matches all of these observations are microfractures that release stress energy generated during cooldown due to differences in thermal contraction between differing materials that compose the detector.  To test this hypothetical production mechanism identical detectors were operated with one structurally supported by low stress wirebonds and the other by a high stress glue joint \cite{Microfractures:Anthony-Petersen:2022}; as shown in the Fig. \ref{fig:tes_spectrum} below. 
Results show the high stress glue joint had nearly a two order of magnitude increase in low energy background rate, strongly supporting the stress relaxation microfracture hypothesis as the culprit of the excess event rate.

For phonon sensitive detector technology to reach its ultimate potential, remaining relaxation event rates must be isolated to a specific material and then mitigated. Additionally, detector technologies must be developed in which any residual relaxation events can be distinguished from putative dark matter interactions.   

\begin{figure}[ht]
    \centerline{
    \includegraphics[width=0.95\columnwidth]{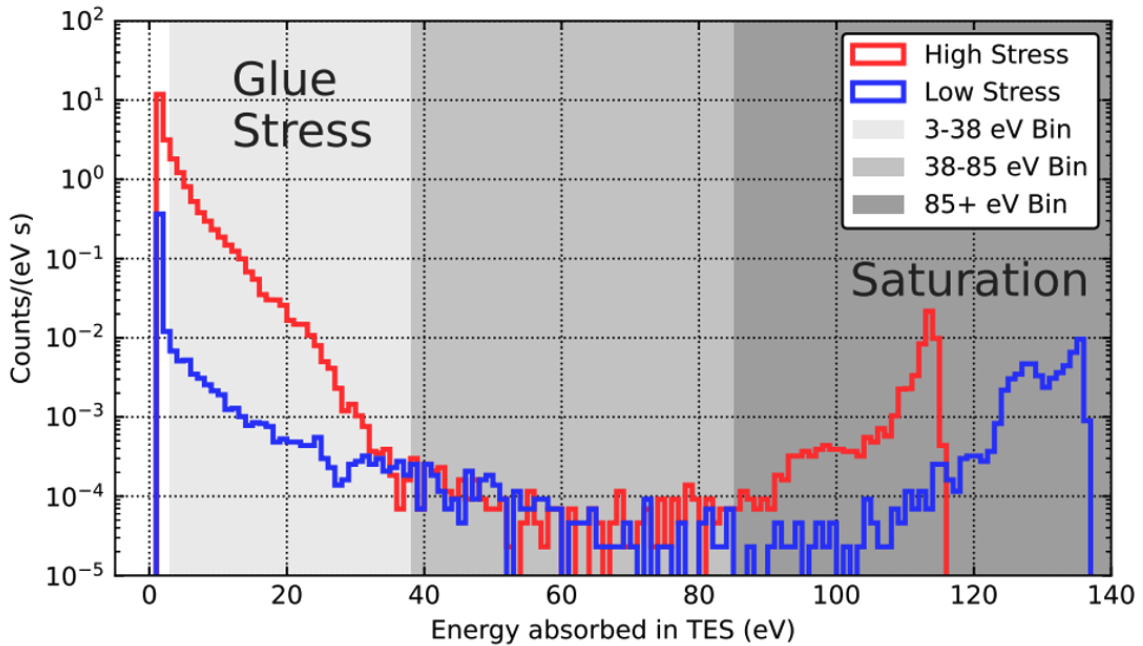}
    }
    \caption{Energy spectrum of energy impulses in a silicon substrate as measured by a transition edge sensor with a low threshold of 1~eV.  The higher energy events are the cosmic rays and ionizing radiation which have caused ``catastrophic" events, wiping out all quantum information in the Google Sycamore chip~\cite{Mcewen2022}.  The lower energy events are consistent with being due to release of mechanical stresses in the devices.  Future studies will push the energy threshold of HEP microcalorimeters even further down to the sub-milli-eV energy scale of single quantum excitations \cite{Microfractures:Anthony-Petersen:2022}.}
    \label{fig:tes_spectrum}
\end{figure}

\subsubsection{Charge Noise}
The phenomenon of charge noise in superconducting qubits led to the development of the charge-insensitive ``transmon qubit" which is the basis of many modern quantum computing platforms.  Two forms include 1) the trapping of charge in defects in the qubit substrate, creating unwanted voltage biases on the qubit via stray capacitive coupling, and 2) liberated charge quasiparticles in the superconductor in the form of broken Cooper pairs which create large qubit frequency jumps as they tunnel across the Josephson junction and change the superselection sector of the superconducting island from even charge parity to odd charge parity or vice-versa.

The charge noise can be viewed as a feature rather than a bug, as liberated charges in otherwise charge-neutral quantum materials is a signature of some energy deposit event.  Understanding the design knobs to turn will enable the engineering of qubits which are maximally insensitive to charge noise as well as of quantum sensors which have maximum quantum efficiency to detect the charge noise.  Understanding the conventional backgrounds using dark matter experimental methodologies will also improve the performance of future quantum computers and future HEP detectors.

\subsubsection{Two-level systems}
Two-level system (TLS) noise in dielectrics is an ubiquitous problem limiting the performance of superconducting qubits, microwave kinetic inductance detectors, and gravitational wave detectors. Progress on this problem would have far-reaching impact in a variety of fields. There are currently significant gaps in our understanding of TLS physics, which can be substantially improved through theoretical and experimental study of materials, including increased connections to the quantum centers and the materials science community.

The TLS can also be viewed as a feature rather than a bug, as these contaminants or defects themselves exhibit single quantum behavior and have a desirable magnetic dipole coupling to electromagnetic fields across a broad spectrum.  While the nitrogen-vacancy center has been deployed in various local sensing applications including of temperature and magnetic fields, various other optically-active color centers have been discovered in common materials used in electronics nanofabrication.  Investigation of the various kinds of TLS will likely lead to benefits in both quantum computing and quantum sensing applications~\cite{Rezai2022}.  

\subsubsection{Materials for spin sensors}
Nuclear and electron spin sensors act as HEP detectors by measuring the dynamics of spin ensembles, as the spins interact with new-physics particles and fields. In order to maximize the science reach of such detectors, it is necessary to optimize the properties of the materials that host the quantum spin ensembles. For example, searches for the EDM interaction of axion dark matter have to use a material where heavy atoms with unpaired nuclear spins are situated at non-centrosymmetric crystal lattice sites. The material properties need to ensure long spin coherence times and the possibility of obtaining large spin polarization~\cite{Sushkov2023b}. Isotopic purification may be necessary to maximize the available spin density and remove unwanted isotopes. Early experiments have focused on ferroelectric crystals, but a broader investment into materials design, fabrication, and characterization can optimize HEP science reach of spin-based sensors.

\subsection{Electronics and integration}
\subsubsection{Room-temperature control and readout electronics (FPGAs and ASICs)}
The main challenges for the control and readout (C\&R) of large scale quantum sensors are maximizing sensor performance while increasing C\&R functionality. Large scale detector C\&R must be able to scale up and minimize cost. The path to that requires:
\begin{itemize}
\item Take full advantage of the frequency and time domains to multiplex sensors, while keeping noise below the sensor noise.
\item Make hardware multichannel and scalable to minimize cost and volume.
\item Make hardware compact by integrating functions as much as possible and clearly defining optimum hardware boundaries (e.g. RF, non-RF, cryogenics, room T, ASIC, discrete, noise critical, high density, etc).
\item Minimize wiring in large sensor array electronics. Eliminate potential sources of noise and crosstalk. Make use of optimum layout design.
\item Develop functionality based on hardware, firmware and low level software optimized to sensors.
\item Develop control system interfaces based on firmware based processors and embedded systems that provide a common platform for a large community of sensors and array sizes.
\item Open source the system to allow sensor developers and experiments to build API interfaces for experiments in a way that fits naturally with the controls.
\end{itemize}

These practices will be enough for mid- to large-scale experiments requiring few million and up to 25 million sensors.  Significant progress in multiplexing the C\&R has already been made with the Quantum Interface Control Kit (QICK)~\cite{Stefanazzi2022}, an open source quantum hardware control platform based on Xilinx RFSOC boards which was developed initially via a QuantISED pilot award and is now being partially supported by the NQISRCs.  This FPGA-based platform is highly flexible and while initially developed on superconducting qubit hardware, is now enjoying widespread adoption in other areas of quantum experimentation.  Importantly, it allows the control and readout strategies to be co-developed with the quantum device hardware so that new and changing electronic needs can be easily implemented in the FPGA. 

For detectors requiring 100s of millions up to a billion sensors (e.g. Windchime) a different approach must be found. For instance, a C\&R system using channel sparsification near the sensor. Sparsification means to promptly detect and only read and process the firing channels. The mid-to-large solution will cost few dollars per channel and fit up to 50K channels per board. In this approach all channels are being read out at all times and channel (i.e. data) sparsification is done at the firmware level. For a billion channels this approach will cost a few billion dollars and occupy 100 full size racks and consume MWs of power. An option for that problem would be based on channel signal detection and sparsification as close as possible to the detector. Channel occupancy is typically low (more so for DM experiments) and sparsification will make compact systems. There are certain challenges to this approach. More complex electronics that allows sparsification needs to be moved near the sensor, inside refrigerators with limited cooling capacity and space. The functionality of that electronics is sensor dependent. For instance, different detectors may be required to measure accurate timing, small charges or powers, single photons etc. 

\subsubsection{Cryogenic electronics (Superconducting electronics, cryoCMOS)}
The same superconducting nanowire technology that enables SNSPDs has recently been used to realize a variety of low-temperature electronics based on devices known as nTrons \cite{McCaughan2014}, including a shift register \cite{FosterShiftRegister2023}, a digital logic family \cite{Buzzi2023}, and a readout for a detector array \cite{Oripov:2023yod}. 
Importantly, this technology is distinct from the SFQ logic family \cite{Ren2022}  
that, while well developed, is unable to operate in even a modest magnetic field.
Some of these circuits have been used for direct discrimination and read out of the nanowire detector signals, and some are targeted at multiplexing to reduce the number of required cables coming out of the cryostat. These technologies are known as Tron-based circuits.

In many situations, cryoCMOS provides a convenient and well-developed, scalable solution for electronics~\cite{Braga2021}.  However, in situations where power dissipation is at a premium, or the electronics are exposed routinely to heavy radiation doses, cryoCMOS may not be suitable.  Thus, a key remaining question about these superconducting electronics is their sensitivity to radiation.  While they are typically fabricated from thicker films that are less radiation sensitive, some errors or damage may still occur, and this effect should be understood before development of large-scale systems can be undertaken. In addition, tools do not yet exist for scaled design and characterization of such a technology.  A key element of the required testing is a platform for control of small digital signals which has recently been developed\cite{FosterThesis2023} based on the QICK platform \cite{Stefanazzi2022}.  

\subsubsection{Cryogenic interconnects}
While superconducting detectors have exquisite sensitivity, a key challenge in scaling to large arrays is the ability to read large numbers of readout lines out of a sub-kelvin cryostat. Even with cryogenic readout electronics, high-density interconnects for superconducting circuits are essential to be able to scale to large sensor formats to maximize HEP science return. Such developments include high-density cable assemblies, integration of the sensors in three dimensions with read-out electronics, and efficient thermalization of large numbers of readout lines. 

\subsubsection{Fast (precision timing) readout}
Though fast and precision timing readout have been successfully demonstrated for single channels, the key challenge is to scalably maintain the excellent performance for a large number of independent sensors and channels. 
The key technical challenges to fully harness the precision timing performance at the detector system level are to develop low power fast (micro-)electronics and to maintain their performances under cryogenic conditions.
While the bulk of these efforts on fast, low power, and cryogenic electronics have been described in the above subsections, one other promising solution is to leverage on the development of high density cables and interconnects to extract the signals from cryogenic to room-temperature conditions where state-of-the-art room temperature electronics are already available to provide existing scalable solutions for fast readout. 

\subsubsection{Common quantum network technology needs for quantum sensor arrays}

Advanced schemes for networked dark matter sensors and for quantum-enhanced telescopy will depend on being able to (i) create and successfully transport more exotic quantum states at various levels of complexity (from W to GHZ), and (ii) collect astronomical photons into single modes and interfere them correctly with generated reference states (see \cite{Khabiboulline_2019a,Khabiboulline_2019b} for example).  Quantum networks involve all of the above aspects and particularly enable longer baselines for improved angular resolutions.  Here we list some technology considerations that apply to long-distance photonic quantum networking generally, and many of which are being addressed at DOE laboratories under QIS portfolios (for matter-matter quantum networks see Section~\ref{AtomicQN}):

\begin{itemize}

\item Quantum networks use entangled light sources to distribute entanglement amongst various nodes. Both discrete and continuous variable network devices are needed.

\item High quality low-loss channels are needed.

\item High nonlinearity materials are needed; integrated sources of nonlinearity are needed.

\item A big gap is the lack of good fiber coupling from chip-scale sources to fiber networks; more work needs to be done to bring this coupling up (to well above 90\%)

\item Repeater technologies are needed. This includes researching quantum memories. However, quantum memories have many expensive requirements and limitations. All optical quantum repeaters would alleviate these issues and save considerable cost. High non-linearities are required for all optical repeaters.

\item Nonlinearities can also be synthesized with photon number resolving detectors. TES devices with near-unit QE and very high count capabilities are needed.

\end{itemize}

\noindent
Multiple DOE labs have extensive experience making entangled light sources. As an example ORNL expertise in both continuous and discrete entanglement can be leveraged. Much work funded by DOE-ASCR in quantum networks research, which often seeks to understand the requirements for functioning quantum repeaters, will also benefit HEP \cite{derevianko2022quantum}.

\subsection{Superconducting device fabrication facilities}

Superconducting devices are a core technology that is critical for enabling a wide variety of the HEP quantum science discussed above. Examples of technologies discussed in this report include: TES, MKID, Qubits, parametric amplifiers, RQUs, SNSPDs, QCD, SQUIDs, Superconducting Digital Circuits, and optomechanical devices for transduction (e.g. coupling to gravity). 

Common needs across a number of the superconducting quantum technologies discussed in this report include: exploration of novel and designer materials to improve and extend device performance, and stable and consistent fabrication of devices and sensors (e.g. low-noise parametric amplifiers, SNSPDs) that are critical to quantum R\&D.  Moreover, integration of these new quantum technologies into fieldable experiments requires: incorporating new materials into robust fabrication processes, optimizing processes to scale to large sensor formats with high yield, integrating devices with readout, integration of compatible processes, and hybridization of incompatible processes. Addressing these needs requires facilities having critical capabilities with materials and fabrication control, 3D integration, and superconducting interconnects. These facilities are critical for HEP as they enable the development and fabrication of essential devices required by the portfolio of new experiments addressing HEP science priorities.

\subsubsection{Existing Facilities and Capabilities}

Superconducting devices and sensors within the HEP program are currently supported through a fairly disconnected collection of individual projects using a variety of cleanrooms. These include small single PI academic facilities and larger shared facilities at national labs. Non-DOE National Lab facilities and FFRDCs, such as MIT/LL, NIST, and JPL, are a valuable asset to the QuantISED program and participate through University Partnerships. In the DOE complex, the Argonne Clean Room (ACR) and the Center for Nanoscale Materials (CNM) at Argonne National Laboratory provide superconducting device fabrication capability. SLAC is commissioning the Detector Microfabrication Facility (DMF), a dedicated facility specialized and optimized for superconducting device fabrication with a focus on quantum devices. The DMF is partially supported by the Q-NEXT DOE QIS Research Center as a Quantum Foundry. Both the DMF at SLAC and the Argonne facilities have the potential to broadly support the HEP community, though operations of these facilities currently are not supported by the HEP program.

As discussed in $\S$\ref{Sec:Strategy}, an important strategy to optimize HEP program utilization of QIS technology for sensing is to create a quantum device fabrication exchange. This structure could be developed to:
\begin{itemize}
\item provide a framework for collaboration between the various facilities maximizing the impact of their collective expertise and equipment
\item  cultivate a coherent program to ensure the broad needs of HEP-specific R\&D are appropriately met with no technical gaps
\item connect groups across the HEP community to these fabrication facilities 
\end{itemize}

\subsubsection{Long-Term Needs for Superconducting Fabrication Capabilities}

As the program develops, there will be a need for R\&D into new fabrication technologies (possibly in collaboration with industry), and the incorporation of new materials into fabrication capabilities. Investment in existing facilities could include not only re-capitalization, to sustain current critical capabilities, but also valuable new tool sets such as atomic layer etching and a broader suite of MEMS, integration, and hybridization tools to expand the range of possible fabrication support for HEP quantum sensor development.

\section{Synergistic activities}

\subsection{Quantum research outside of HEP}
Quantum information science is currently a high priority for the U.S. in order to maintain the country's leadership and economic competitiveness in current and future quantum technologies.  Large investments have been made through various governmental funding agencies including DOE, DOD, NSF, and NIST, and there are currently efforts to expand the quantum funding umbrella to other agencies including NASA.  The 2023 NASA Biological and Physical Sciences decadal survey~\cite{NAP26750} included “\textit{What new physics, including particle physics, general relativity, and quantum mechanics, can be discovered with experiments that can only be carried out in space?}” as one of the key questions and  proposed a multiagency  “Probing the Fabric of Space Time (PFaST)” initiative with quantum sensors in space for fundamental physics studies.

Due to the nature of quantum computing in which individual quanta of information are manipulated to gain quantum advantage through coherent processing, the elements of quantum computing are intrinsically sensitive to small disturbances at the single quantum level.  Ongoing investments in QIS in developing single quantum devices and manipulating this information can thus be efficiently leveraged to provide low energy threshold, single quantum sensing for HEP science applications. 

The field of quantum computing has been careful not to over-promise on near term capabilities as universal quantum computing may still be technologically a decade away.  A strategy that has emerged for that field is to focus on what can be done with near-term, ``noisy intermediate-scale quantum" (NISQ) technologies.  One of the key applications is quantum sensing, by which a steady stream of good news in fundamental science results emerging from quantum technologies would keep public attention focused on QIS research.  It has been remarked by a leading QIS experimentalist that quantum devices are only good for sensing things that can penetrate the cryostats, radiation shields, Faraday cages, and vacuum systems of the experimental apparatus -- elements that are meant to protect the fragile quantum information from environmental disturbances.  Other than novel cryogenic quantum materials that can live inside the cryostats, HEP thus provides most of the key science targets for quantum sensing.  These come in the form of weakly interacting BSM particles and their associated forces which are far weaker than electromagnetism, and thus allow these new particles and waves to easily penetrate through electromagnetic shielding techniques employed in quantum computing platforms.  New and unprobed cosmological phenomena are also potentially observable through gravity which also cannot be shielded against.  

Close collaboration with leading scientists outside of HEP could therefore provide useful benefits not only to HEP, but also to the broader QIS community.  It should however be recognized that while quantum sensor development possibly including pathfinder and proof-of-principle experiments should necessarily be a collaborative and cross-disciplinary effort, it would be unrealistic to expect non-HEP agencies to support full-scale projects targeting primarily HEP science.  

\subsection{Quantum Detector R\&D coordination effort in the US and collaboration with Europe}
In the US, the Coordinating Panel for Advanced Detectors (CPAD) formed in March 2023 a new network of US Detector R\&D Collaborations including groups targeting Low Background Detectors (RDC7) and Quantum and Superconducting Sensors (RDC8).  The  goal of these RDCs is to cover major technology areas aligned with the 2019 HEP Basic Research Needs Workshop on Detector R\&D~\cite{DetectorBRN2020}, with Snowmass and P5, and with the 2023 Quantum Sensors for HEP workshop described in this report. The goal is to bring together the community in a more persistent way than the annual CPAD workshops alone, to coordinate R\&D efforts and to forge collaborations. Similarly, the quantum information science effort in Europe has been primarily outlined through the European Committee for Future Accelerators (ECFA) Detector R\&D for quantum sensing for particle physics (DRDQ/DRD5) program with a white paper planned for 2024.

While the activities of the present Quantum Sensors for HEP workshop occurred prior to more recent meetings under the auspices of CPAD and ECFA, it would make sense to coordinate quantum detector development strategies with these groups, especially as large collaborations will be necessary to scale up quantum sensing efforts to full HEP experiments.  Two participants in the present workshop have volunteered and are now co-leading two of the CPAD RDCs, Noah Kurinsky for RDC7 and Rakshya Khatiwada for RDC8.  The CPAD groups also plan to have regular communications with their ECFA counterparts.

\section{DOE capabilities and expertise}
National labs and university groups funded by DOE bring new capabilities to the table to advance QIS research.  The first step in this direction was the funding of 5 National Quantum Initiative Science Research Centers, each with a DOE national laboratory serving as the lead lab.  These include SQMS (FNAL), QSC (ORNL, FNAL, LANL), Q-NEXT (ANL, SLAC), C2QA (BNL), and QSA (SNL, LBNL).  Quantum sensing comes as a natural application of this science and technology development. However, specific sensor needs are different in HEP applications, so sensor development for HEP science will need to be the under the purview of HEP programs.

Techniques established in underground dark matter science including mitigation of cosmic rays, muon hodoscope vetoes, selection of radiopure materials, and calibrated particle beams can be brought to bear on engineering quantum systems with far reduced background rates.  Once demonstrated, these techniques can be transferred to the quantum computing field.  Similarly, HEP capabilities in controlling hundreds of millions of electronics channels in complex particle collider detectors and astronomical sensor arrays can efficiently be transferred to the control of large arrays of quantum sensors, and eventually to the control of million-qubit quantum computers.  With this HEP know-how, the large scale quantum computers of the future could perhaps themselves be operated as quantum sensor arrays, simply by monitoring their qubit error rate.

A list of unique DOE capabilities relevant to quantum sensor development and to the deployment of quantum sensors for HEP science experiments includes the following.
\begin{enumerate}

\item Project management, integration experience for large scale projects.  
A unique asset of DOE is the ability of its national labs to manage and conduct full scale HEP experiments with discovery sensitivity.  While small scale experiments utilizing quantum sensors can be conducted in small research labs and funded as proof-of-principle demonstrators from HEP or non-HEP sources, ultimately the full-scale deployment of quantum sensors in HEP experiments will benefit from DOE national laboratory project management expertise. 

\item Large-scale applied physics, engineering, and technical resources (mechanical, electrical, cryogenic).  These can support quantum sensor R\&D efforts with scalable solutions that can be implemented in future intermediate and large-scale systems.  These include:
\begin{itemize}
\item Accelerator technologies expertise:  Cavity design, testing, and operations expertise, especially for SRF cavities which exhibit record lifetimes. These resonators enable coherent accumulation of tiny signals while providing narrow bandwidth for noise rejection.  The long lifetime results from isolation from lossy environments, and thus also enables high fidelity storage of fragile, non-classical states of light which may be used to enhance sensitivity to rare processes. DOE accelerator labs also have ultra-high-vacuum engineering and operations capabilities which will be similarly important for reducing decoherence and heating in quantum sensors based on atoms and ions.

\item Superconducting magnet design, testing, and operations expertise, enabling the large scale magnet systems needed for axion dark matter experiments and other probes of CP-violation.  HEP magnet test stands can be used to develop quantum sensors which are compatible with a magnetic field environment.

\item Mechanical and cryogenic engineering to design, build, and implement unique detector prototypes meeting the unique requirements of HEP experimental settings.  These engineering teams will also be crucial for scaling up the quantum sensing platforms to full-scale HEP experiments.

\end{itemize}

\item{Instrumentation expertise -- HEP possesses vast and unique instrumentation expertise from fielding large, complex detector systems which are operated in unusual and often harsh environments required by HEP science.  This expertise is one of the key advantages HEP brings to the table in the quantum sensor co-design effort.   Key capabilities include the following.}
\begin{itemize}
\item Cosmic detector expertise:  Superconducting sensor arrays and Josephson junction circuits, including TES's, SQUID multiplexers, and microwave SQUID upconverter multiplexers have been pioneered by scientists in the DOE complex for applications in dark-matter detection and cosmology. These superconducting sensors are being generalized to quantum sensing implementations, including the first deployment of superconducting qubits as single photon detectors for HEP applications.

\item Low-background expertise and facilities, including underground test facilities at FNAL, PNNL, and SNOLAB.  Production and handling of radio-isotope sources.  HEP dark matter experimenters have a wealth of experience in understanding and controlling the backgrounds in particle detectors, and this experience can be applied to understanding the environmental disturbances affecting ultra-low threshold quantum sensors including the various types of solid-state qubits.

\item Large-scale HEP electronics instrumentation expertise including the scalable QICK FPGA control system, EPICs for real time experimental control, and microelectronics departments for custom ASICs.  This instrumentation capability is needed for fielding large arrays of quantum sensors and other complex hardware systems.  HEP instrumentation is already providing quantum rf control for state preparation and multiplexing of systems of quantum devices, and cryoCMOS quantum control electronics are also being developed.

\item  Customized electronics for operating interferometers, feedback/control of lasers for high-speed, high bandwidth noise suppression.  Integration. Laser noise suppression + Interferometer + advanced readout filter system + state preparation + data analysis.  As interferometers and other quantum sensing systems become more complex, a lab-scale effort will be required to streamline the hardware operations.

\item Telescope optics and focal plane instrumentation.  BNL has a lab that is currently studying the integration of fast timing single photon detectors into telescopes, and prototyping (i) fast spectrograph based on SPAD arrays; (ii) telescope-to-single mode coupling and long-distance quantum state transport.  In addition, existing materials-processing facilities at DOE labs (e.g. BNL, Argonne) could be employed in fabrication of custom SNSPDs for this application.

\end{itemize}

\item{Unique infrastructure from decades of investment in HEP labs and projects}
\begin{itemize}
\item Deep shafts/underground sites.
In addition to the low background test infrastructure, the deep underground sites operated by DOE provide new and unique opportunities for low background experiments and for atom interferometers probing long wavelength phenomena and requiring large free-fall times.
\item HEP detector development infrastructure, including cryogenics handling, test beam and irradiation facilities e.g. at FNAL and SLAC, and supporting readout electronics.  These unique facilities may be used to engineer radiation hard quantum electronics for high intensity beam experiments -- yet another non-traditional environment for quantum sensors.
\item High-performance computing for large-scale data analysis.  As quantum devices and sensor arrays mature, they will likely produce an even larger flood of data than even the most highly multiplexed classical HEP detector systems operating today.  Future quantum experiments will likely need hybrid classical-quantum control systems along with real-time, in-situ data processing.

\item Large cryogenic plants with technical staff for operations.  These lab-based resources support a fleet of cryogenic test stands needed for the development of microcalorimeters, single photon detectors and other quantum sensors.  They also offer the infrastructure needed for large scale deployment of quantum sensors in an experimental setting.

\end{itemize}

\item{Theory groups closely integrated with experimental efforts}
\begin{itemize}
\item Strong particle physics theory support, now including exploration of new HEP experimental techniques utilizing quantum sensors developed in other fields including AMO. 
\item Deep expertise in extracting science from experimental data, including for example integrating multi-modal/multi-messenger observations into cosmology models. 
\end{itemize}

\item Resources, capabilities, and infrastructure funded from programs outside of HEP are also available within the DOE national laboratory complex and are potentially accessible for quantum sensor development.  Some of these external investments are already being leveraged to benefit the HEP program.  For example, the large investment by ASCR in quantum networking provides the infrastructure to co-develop back-action evading measurements that could be expanded to operate over large quantum sensor networks.  Some of the available resources being supported by other programs include:
\begin{itemize}

\item Ultrastable fiber and free space networks, including classical networks of quantum sensors such as those of atomic clocks.  There are also a few new quantum network testbeds in the DOE labs. These can be tailored towards HEP science, for example using entanglement distribution for quantum sensing of extremely long coherence length signals.  Extensive experience making entangled light sources including ORNL expertise in both continuous and discrete entanglement can be leveraged.  Much work funded by ASCR in quantum networks research, which often seeks to understand the requirements for functioning quantum repeaters.  

\item Quantum optics labs that can surpass the SQL in optomechanics/MEMS.  Multiple labs specialize in transduction which will be critical for quantum networks of optomech devices, and especially for coupling them to other quantum devices like transmons.  

\item Device fabrication facilities at SLAC DMF, ANL ACR, Sandia, (+User facilities within DOE:, ANL CNM, LBNL).  To meet HEP device needs, engagement is also needed with key foundries outside of DOE including NIST, JPL, MIT/LL, NASA Goddard, Pritzker Nanofab, possibly industry fab.  These will be critical for fabricating superconducting devices, optomechanical devices, transduction devices, that will be useful for a variety of sensing applications.

\item DOE quantum computing testbed facilities (e.g. AQT, SQMS).  Can potentially be used to study algorithmic processing of data from quantum sensor arrays in order to identify signal events and reject environmental background disturbances.  

\item DOE materials science centers including CINT at LANL and CNMS at ORNL provide crystal growth and quantum materials synthesis and characterization capabilities.  Also LBL quantum foundry, and cryogenic test capability at SLAC, FNAL.  Crystal growth will be the main limitation to scaling to experimental scale, and assay is needed for the source material for all of these new materials.

\item Accelerator facilities, e.g. FRIB, for production of interesting isotopes to produce sensitive molecules for AMO sensing platforms.

\end{itemize}
\end{enumerate}

\section{Personnel and workforce development}
While the workshop discussions did not specifically address workforce development, it can be noted that at present, there are limited opportunities for training and workforce development specifically for the application of quantum information methods and techniques in HEP.  The establishment of a stable QuantISED-like program would address many of the challenges facing this cross-disciplinary research by overcoming the silo-ing of academic disciplines into ``high-energy" and ``quantum."  
The availability of a long-term, reliable funding source is a key factor in workforce preservation and development in all fields, not just in QIS.  Funding gaps, lack of certainty about funding renewal, and large time investments for repeated grant writing due to limited program funding serve to disincentivize individuals from remaining in the field.  While these concerns are ubiquitous, in QIS they are felt particularly because such attractive alternative options exist in industrial settings.

Other possibilities for HEP to help build a strong quantum workforce include providing training grants to support research assistantships and postdocs at universities and national laboratories and CAREER/ECRP-type grants to support junior faculty interested in pursuing this area.

\newpage

\begin{appendices}
    
\section{Charge\label{s:charge}}
Dear Colleagues, \\
\\
Ultra-precise and -sensitive measurements of quantum phenomena offer opportunities to
devise powerful probes to explore physics relevant to the mission of particle physics. By
leveraging and driving advances in quantum sensing, it may be possible to explore a broader
range of science parameters than would otherwise be achieved, opening the possibility to
search for novel fundamental particles and their interactions beyond the Standard Model of
Particle Physics (e.g., searches for dark matter, dark energy surveys, precision studies of
fundamental constants, anomalous spatial variations or oscillations of physical interactions,
and other phenomena).
We request that you plan a workshop to review status and plans for Quantum Sensing for
High Energy Physics (HEP). To explore and adopt QIS technologies and tie them to future
particle physics experiments is a key goal of this workshop. It follows from and builds on
previous workshops and planning exercises, including the Quantum Sensing for HEP
workshop at Argonne National Laboratory in Dec 2017~\cite{ahmed2018quantum}, the HEP Instrumentation Basic
Research Needs workshop in Dec 2019~\cite{DetectorBRN2020}, and the HEP QuantISED sensors workshop in
May 2020~\cite{barry2021opportunities}.
The workshop should identify particularly promising approaches in the domain of quantum
sensing that can be utilized by future HEP applications to further the scientific goals of the
community as outlined by the P5 Report~\cite{P5report2014} and in the recent Snowmass 2021 Report~\cite{osti_1922503}.
Gaps in the current R\&D strategy that could impact the science goals should be identified.
Synergistic activities such as development of relevant quantum algorithms and protocols, or
simulation tools in support of quantum sensor development, would be within the scope of the
workshop.
The techniques used in Quantum Information Science span a broad spectrum and many are
not familiar to the HEP community. As there is substantial experimental expertise across
several fields—much of it outside the particle physics community—this workshop should
reach out to invite scientists with relevant expertise in quantum sensing to help guide the
HEP program in utilizing QIS technology.
We anticipate the workshop would benefit from a Scientific Program Committee to review
and refine the scope of the material covered at the workshop, and to suggest the most
appropriate speakers to deliver that program.
We appreciate your willingness to take on this important task. We would appreciate a
summary of the important findings and comments from the workshop by Jun 30, 2023, and
a final report by Sep 30, 2023.\\

\noindent Sincerely,\\
\\
\noindent Glen Crawford\\
Director, Research and Technology Division\\
Office of High Energy Physics\\

\section{The DOE HEP Quantum Information Science Enabled Discovery program (QuantISED)}
\label{a:quantised}
The QuantISED program grew out of the nascent DOE HEP Connections program which was itself motivated by the 2014 P5 report~\cite{P5report2014} which highlighted cross-disciplinary connections as a way to bring new technologies to bear on HEP science.  QuantISED was launched by a roundtable meeting held in February 2016 which resulted in a report titled ``Quantum Sensors at the Intersections of Fundamental Science, Quantum Information Science, and Computing"~\cite{osti_1358078} which identified various use cases for quantum information science and technologies to be potentially applicable to HEP science targets.  The QuantISED program was launched in fiscal year 2018 with research programs targeting quantum theory, quantum algorithms, quantum hardware, and quantum sensors.  Its current portfolio includes ongoing R\&D in quantum sensors targeting HEP science as well as two small experiments.  The fixed term funding supports both HEP scientists and engineers as well as key researchers in other disciplines which have not traditionally been funded by HEP.  This program is summarized at \href{https://science.osti.gov/hep/Research/Quantum-Information-Science-QIS}{https://science.osti.gov/hep/Research/Quantum-Information-Science-QIS}.

\section{Experimental concepts presented at the town hall}
\label{a:townhall}
The slides from the town hall presentations are available at the workshop website:\\
\href{https://indico.fnal.gov/event/59102/}{https://indico.fnal.gov/event/59102/}.
\begin{enumerate}
\item Aaron Chou (FNAL),
The need for quantum sensors for HEP science.  \\ Key point:  Quantum sensors with sub-eV energy thresholds enable new probes of nature which are not possible with current eV threshold detector technologies.

\item Matt Shaw (JPL),
Superconducting Nanowire Single Photon Detectors with Ultra-low energy threshold.  \\ Key point: SNSPDs have reached a new record low threshold, counting single far infrared photons at 43~meV or 10~THz.  

\item Cristián Peña (FNAL),
Axion DM with low-threshold SNSPDs. \\  Key point: low threshold SNSPDs have a dark count rate of $10^{-5}$~Hz and could be used in broadband axion searches such as BREAD.

\item Stewart Koppell (MIT),
Improvements to the LAMPOST Experiment. \\  Key point:  The LAMPPOST hidden photon search is being upgraded with imaging superconducting nanowire sensors which will provide spectral, spatial, and amplitude resolution.

\item Si Xie (FNAL),
Precision Timing and Scalable Readout for low threshold SNSPDs. \\  Key point:  The Fermilab QICK board can be used for fast and scalable readout of large arrays of SNSPDs, using frequency domain multiplexing.

\item Stewart Koppell (MIT),
Dielectric Powder as an Axion/Dark Photon Haloscope. \\  Key point:  Space-filling 3D microstructures can be used to maximize the scattering surface area in broadband wave dark matter experiments.

\item Pierre Echternach (JPL),
Quantum Capacitance Detectors for Terahertz Single Photon Counting. \\  Key point: Charge qubits can be used as single THz photon detectors with low dark count rate, targeting dark radiation and dark matter in the form of axions and dark photons.

\item Rakshya Khatiwada (FNAL),
Dark Matter detection with Quantum Capacitance Detectors. \\  Key point: Far infrared single photon detectors are needed for dish antenna experiments like BREAD, but they also need to be paired with FIR calibration sources which are under development.  

\item Ritoban Basu Thakur (JPL),
Kinetic Inductance Traveling-Wave Parametric Amplifiers. \\   Key point:  Kinetic inductance in thin film superconductors can be used to create nonlinear devices such as parametric amplifiers, mixers, delay lines, and integrated devices such as the Superconducting On-chip Fourier Transform Spectrometer.

\item Yogesh Patil (Yale),
Cavity Optomechanical Search for Axions. \\  Key point:  Single photon detection can be used for single phonon detection in superfluid helium-filled cavities, enabling tests of certain models of quantum gravity. 

\item Lee McCuller (Caltech),
Converting Interferometers into HEP dectors with high-isolation single-photon detection. \\  Key point:  Single photon counters can be used as the readout of optical interferometers to reduce noise far below the standard quantum limit, vastly outperforming squeezing\cite{mcculler2022singlephoton}. Doing so can be viewed as quantum enhancement, but also opens additional possibilities to use state preparation to further enhance sensitivity.

\item Holger Mueller (Berkeley),
Testing the standard model and probing the dark sector by measuring the fine structure constant. \\  Key point: Next generation atom interferometry experiments are poised to achieve another order of magnitude in sensitivity in measurements of the fine structure constant.

\item Sanha Cheong (Stanford),
MAGIS: Extending High Energy Physics with Atom Interferometry. \\  Key point:  MAGIS-100 is deploying a 100~m atom interferometer in a Fermilab underground access shaft in an experiment targeting ultralight dark matter, mid-band gravitational waves from cosmological sources, and tests of quantum mechanics at an unprecedented scale. 

\item Juli\'{a}n Mart\'{i}nez-Rinc\'{o}n (BNL),
Distributed Atomic Sensing in the Long Island Quantum Network.  \\ Key point: With funding from New York state, Stony Brook and Brookhaven are deploying a 5-node, 240~km quantum network in which entanglement distribution to distant atomic clouds will enable quantum sensing of long-distance, long-wavelength phenomena.

\item Yu-Dai Tsai (UC Irvine),
Direct detection of ultralight dark matter with space quantum sensors.  \\ Key point:  Satellite-based atomic clock and precision tracking experiments will provide unprecedented sensitivity to long wavelength ultralight dark matter, the cosmic neutrino background, and hidden fifth forces.

\item Paul Stankus (BNL),
Quantum-Assisted Optical Interferometry for Precision Astrometry.  \\ Key point: Quantum engineering can improve astronomical interferometry, including Hanbury-Brown-Twiss experiments, for high-resolution imaging and precision astrometry, targeting HEP cosmology goals including Hubble tension, dark energy, dark matter, and pre-CMB relics, as well as tests of general relativity.

\item Peter Cameron 
Training chatGPT on quantum impedance networks of QED.  \\ Key point: ChatGPT points out interesting connections between seemingly disparate sectors of physics, pointing towards poor impedance matching as the cause of weak coupling to dark sectors.

\item Francois Leonard (SNL),
Exploiting the Physics of the Field for Compute-in-Sensor.  \\ Key point:  The analog nature of wave signals can be exploited by patterning quantum sensors to perform pattern recognition, pre-filtering, and frequency selection in order to reduce time and energy demands on digitization and pre-processing.

\item Matt Pyle (Berkeley),
Understanding and Mitigating quasi-particle excess due to phonons and IR produced by stress relaxation.  \\ Key point:  Energy relaxation events such as stress-induced microfractures in solid state detectors may be responsible for the excess quasiparticle density seen in superconducting devices, and understanding and mitigating this background will be crucial to both future dark matter searches and quantum computers.

\item Sergey Pereverzev (LLNL),
Material science of quantum sensors.  \\ Key point:  Materials-induced low energy backgrounds have been ubiquitous in dark matter searches, and understanding and mitigation of these backgrounds will be crucial for low threshold quantum sensing.

\item Rakshya Khatiwada (IIT/FNAL),
Sapphire substrate qubits for low mass Dark Matter searches.  \\ Key point: Superconducting qubits can be used to read out single phonon excitations in sapphire and other low phonon gap crystals used as dark matter scattering targets.

\item Alexander Sushkov (Boston University),
Quantum engineering of macroscopic spin ensembles to search for QCD axion dark matter. \\ Key point: Nuclear spin ensembles are used to search for the defining EDM interaction of the QCD axion, over a uniquely broad six-decade range of masses, corresponding to frequencies between hundreds of Hz and hundreds of MHz.

\item Dylan Temples (FNAL),
Optical Strain Sensing for Particle Detection.  \\ Key point:  Optical strain sensors provide another way to measure phonon signals in solid-state dark matter targets and provide various measures of the dark matter interaction including possible electron/nuclear recoil discrimation, background rejection of microfracture events, and searches for resonant lattice scattering processes with no phonon signal.

\item Javier Tiffenberg (FNAL),
Closing the Loop on Quantum Research with Skipper-CCDs: DOE-OHEP's Contribution to Advancements in Quantum Sensing.  \\
Key point:  The single electron detection capabilities of the Skipper-CCD have enabled new quantum sensing capabilities including an infrared photon-number-resolving imager, qudit tomography, and efforts to build a quantum infrared spectrograph -- a quantum radar using IR photons.

\item Daniel Carney (LBNL),
Nuclear decays with mechanical quantum sensors.  \\ Key point: Nanometer-scale, levitated sensors with embedded radioisotopes can be used to measure the kinematics of nuclear decays to search for emission of BSM particles such as heavy sterile neutrinos or axions as well as to perform precision electroweak measurements.

\item Rafael Lang (Purdue),
The Windchime Project.  \\ Key point:  Arrays of accelerometers with quantum-enhanced readout are being developed to search for passing dark matter via the long range forces exerted on the test masses.

\item Claire Marvinney (ORNL),
Back action evasion and quantum noise reduction in quantum magnetometers for particle and field detectors.  \\ Key point:  Piezo-magnetic materials can be used for quantum non-demolition measurements of the motion amplitude of accelerometers, thus evading the standard quantum limit back-action of direct measurements of position and momentum.

\item Alberto Marino (ORNL),
Quantum Enhanced Detection of Quantum Fields and Particles through Networked Entangled Sensors. \\ Key point:  A distributed array or network of entangled optomechanical sensors is being developed for force detection, and will enable searches for dark matter and BSM physics.

\item Dalziel Wilson (U. Arizona),
Entanglement-enhanced optomechanical dark matter detectors.  \\ Key point:  Distributed entanglement in an array of M quantum sensors will provide $M^2$ scaling in the signal-to-noise ratio, far better than that achieved by classical averaging.

\item Reina Maruyama (Yale),
Rydberg atoms as single-photon detectors for axions.  \\ Key point:  Resonant transitions in Rydberg atoms can be used for single photon counting, thus reducing noise far below the standard quantum limit for axion searches in the 10~GHz frequency range.

\item Itay Bloch (LBNL),
Noble and Alkali Spin Detectors for Ultralight Coherent darK matter (NASDUCK)  \\ Key point:  Quantum spin sensors are necessary for measuring the couplings of axion-like particles to standard model fermions in Noble and Alkali target materials.

\end{enumerate}

\section{Group discussion questions posed to workshop participants
\label{s:questions}}
In order to collect information about science and technology ideas currently being explored by the cross-disciplinary quantum sensors community, the following questions were posed to the workshop participants.
\begin{enumerate}
\item What is the science target and how is it mapped onto current DOE priorities?
\item What are the new quantum technologies and how are they crucial to enable HEP science reach beyond current technical capabilities?  Does the strategy leverage investments in quantum technology from other funding sources?  Are there dependencies on quantum technology development happening outside of the scope of DOE-OHEP?
\item What DOE expertise/facilities/staff/capabilities (not just money) can be usefully brought to bear?  What are current funding sources?  
\item For informational purposes only, regarding potential future experiments:  What is the time scale to a shovel-ready deployment?  What are the key metrics that still need to be demonstrated?  What is the projected budget and scope?  E.g. larger project ($> \$20$M), or small ($< \$20$M) project?  How many institutions/PIs in collaboration?
\end{enumerate}

\section{Workshop organizing committee}
Aaron Chou, Fermilab \\
Kent Irwin, Stanford University and SLAC \\
Reina Maruyama, Yale University (local organizer) \\
Kathryn Zurek, Caltech (theory advisor)

\end{appendices}

\bibliographystyle{elsarticle-num} 
\bibliography{qis_hep.bib}

\end{document}